\documentstyle[12pt,epsfig,graphicx]{article}
\begin{document}
\title{Empirical formulae to describe some physical properties of small groups of protogalaxies with multiplicity}

\author{Guillermo Arreaga-Garc\'{\i}a \\
Departamento de Investigaci\'on en F\'{\i}sica, Universidad de Sonora. \\
Apdo. Postal 14740, C.P. 83000, Hermosillo, Sonora, Mexico. \\
{\it guillermo.arreaga@unison.mx}\\
}
\maketitle
\begin{abstract}
By means of identical cubic elements, we generate a partition of a volume in which 
a particle-based cosmological simulation is carried out. In each cubic element, we determine the 
gas particles with a normalized density greater than an arbitrarily chosen density threshold. By using 
a proximity parameter, we calculate the neighboring cubic elements and generate a list of 
neighbors. By imposing dynamic conditions on the gas particles, we identify gas clumps and 
their neighbors, so that we calculate and fit some properties of the groups so identified, including the 
mass, size and velocity dispersion, in terms of their multiplicity 
(here defined simply as the number of member galaxies). Finally, we 
report the value of the ratio of kinetic energy to gravitational energy of such dense gas 
clumps, which will be useful as initial conditions in simulations of gravitational collapse of 
gas clouds and clusters of gas clouds.
\end{abstract}

{\it keywords:--galaxies: kinematics and dynamics;--physical processes:hydrodynamics;--methods:numerical}
%%%%%%%%%%%%%%%%%%%%%%%%%%%%%%%%%%%%%%%%%%%%%%%
\section{Introduction}
\label{sec:int}

The gravitational collapse of gas clouds takes place by the accretion of gas from low-density to 
high-density regions, such that the result of this process can be the formation of galaxies or 
stars, depending on the density and length scales involved, see \cite{klessen}. In an intermediate stage of the 
collapse process, it is possible that these dense gas structures show a dynamic behavior characteristic of 
the final systems already formed, be they galaxies or stars. For this reason, those intermediate structures 
are called proto-galaxies or proto-stars. In both cases, they can be identified with very dense 
gas cores, which do not shine on their own.

The idea about the existence of a direct relationship between the proto-structures and their 
descendants, whether galaxies or stars, will have more support as long as the collapse of the gas cores 
occurs in a more or less isolated way, with each one producing one or a few final products, see 
\cite{PadoanNordlund}, \cite{HennebelleChabriera}, \cite{HennebelleChabrierb}, \cite{Oey} and 
\cite{Hopkins}. It is in this last point, on which the attacks of the 
opponents to this idea are based, because they argue that stochastic processes are inherent to 
the general process of formation of structures, see \cite{Bonnell}, \cite{Bate} and \cite{Clark}. 

A remarkable example of this idea is seen in the similarity between the core mass function (CMF) and 
the stellar mass function (IMF), see \cite{offner}. Both functions share the same mathematical 
form, with the only difference between them being that the CMF is shifted to larger masses with respect to the 
IMF. 

It is an observational fact that galaxies tend to agglomerate in bounded 
structures, which are called depending on the number of members, for instance 
clusters, with a number of galaxies varing within 30-300; loose groups 
with 3-30 members; compact groups with 4-8 and binaries with 2 members. In 
addition, 30 percent of the galaxies are observed to be isolated, 10 percent forming 
binaries; 0.1 percent in compact groups; 50 percent of the galaxies are observed in 
loose groups while only 10 percent is forming clusters, see \cite{mamon}. In addition,  
it is now well-known that about 50 percent of all galaxies in the Universe are collected into 
low-multiplicity groups, with four or less member galaxies. Observationally, \cite{hcg}, reported 
the properties of one hundred compact groups of galaxies identified in the Palomar Observatory Survey. 
Consequently, one may suspect that the gas clumps out of which galaxies form by gravitational 
collapse, are perhaps also grouped in small groups with a few members. If such a relation exists, then 
it maybe can be seen in cosmological numerical simulations.

With regard to numerical aspects, there are 
many problems to identify a galaxy within the framework of a 
cosmological simulation to compare their clustering properties with observations. Characterizing the 
clustering of matter in a cosmological 
simulation is a complicated issue that has been considered over the last 30 years. Many codes 
have been developed and tested to analyze the highly irregular and filamentary 
clumpy structure of the simulations. A summary of the development of this area has recently been 
presented by \cite{halosmad}, who compared the results obtained by using some dark-matter halo 
finder codes on the same test data with either a cosmological simulation or a mock catalog of dark-matter halos. 

In general terms, the basic objective of most of these codes has been to identify 
isolated dark-matter halos, see \cite{hop}. Most of the early codes discarded the clustering of gas in 
their calculations, partly (i) because they were applied to dark-matter only simulations or (ii)  
because of the well-established idea that dark-matter was clustered earlier and shortly after, the gas reached 
the center of these dark-matter structures to form dense gas clumps, see \cite{white}. Recently, a new generation of 
more refined codes has focused on the determination of sub-halos embedded within a larger host halo, which is a 
harder computational problem; see for instance \cite{canas}. It should be noted that many of these codes are not 
publicly available. Other codes that are open source, can be difficult to understand and run because a lot of parameters 
are involved. 

In this paper we consider the gas component of a typical hydrodynamical cosmological simulation, which tries 
to imitate the Illustris simulation, that was described by \cite{illustris}. The 
size of the simulation box (around 106 Mpc), the values chosen for the content of matter, the expansion 
rate H$_0$ of the universe and other parameters, that were used in the Illustris 
simulation, have also been used in the lower resolution simulation presented in this paper, which 
was developed with the publicly available code Gadget2; see \cite{springel}. This set of 
cosmological parameters has been determined using the most recent
observations, see \cite{planck2014}, and are currently one of the most accurate. It must be emphasized that the simulation 
used for this paper should not be compared to the Illustris one, because adopting the same cosmological parameters does 
not make them equivalent. Important features such as star formation, cooling, feedback, etc., which were included 
in the Illustris simulation are not included in the simulation used in this manuscript.  

We next apply our mesh-based code to generate a partition of the 
simulation box in terms of identical cubic elements, at the scale of 0.8-1.6Mpc. We then 
determine a subset of cubic elements, whose average normalized density above a threshold value is given 
in advantage, at the order 30-300 times the average cosmic density. We then considered the densest cubic 
elements to identify isolated gas 
clumps and produce a list a neighboring gas clumps. Next, we counted the number of groups 
detected in terms of their multiplicity (the numbers of members or richness) and we calculated the physical 
properties of these groups in a statistical sense, including the mass, size, velocity dispersion and 
multiplicity function of the gas clumps. 

The particular partition sizes and overdensities considered in this paper as free parameters, can be 
motivated as follows. The spherical top-hat 
approximation considers that an overdense matter expands with the Universe up to 
the turnaround point, where it stops expanding and starts collapsing. Later, the 
surrounding regions will follow its collapse gradually. The turnaround radius 
of cosmological structures in the top-hat approximation varies from 
1.7 Mpc for small galaxy groups to 7 Mpc for large galaxy clusters. In addition, cosmological N-body simulations 
have confirmed the top-hat approximation 
since a long time ago, so that the radius in which the overdensity is 178 times 
the mean density determined the limits of the infalling region.     

It must be clarified that the computational method described above is not 
new to the field. Consequently, there are no advantages of this algorithm over those in the 
literature. For this, it should be considered as a less sophisticated
version of the mentioned above methods, see \cite{halosmad}. \cite{dobbs} considered a clump-finding algorithm 
based only on a surface density threshold criterion, so that a galaxy model is divided into a Cartesian grid; this 
clump-finding algorithm allowed them to study the evolution of giant molecular clouds. Motivated by this results, we apply 
our code at its first stage of development, since there is no need to 
refine it any further to achieve the objectives of interest, which we outline 
as follows. In addition, in cosmological simulations, the most commonly used value of 
overdensity to define a dense clump is 200 times the average cosmic density, see for instance \cite{halosmad}.     

First, to study the dynamics of gas at sufficiently large scales, 
in order to characterize its grouping properties. Second, to calculate the ratio of kinetic energy to gravitational 
energy of the gas clumps. Therefore, the physical 
properties of giant gas complexes (typical radius, 
mass, 3D-velocity dispersion and the level of kinetic energy) obtained as results from 
the calculations reported in this paper are expected to be relevant as a suitable sets of 
initial conditions from cosmological simulations, to address the problem of the gravitational 
collapse of clouds and clusters of clouds. 

Several comments are now presented to emphasize some potential benefits of this paper. First, that 
this problem of the gravitational collapse of clouds and clusters of clouds can not be studied directly 
by means of present-day cosmological simulations, because the resolution is insufficient to resolve 
properly the scales needed. To alleviate this situation, the zoom-in simulations were introduced since a 
long time ago. For example, \cite{sugi} smoothed 
the initial distribution of particles and obtained a smoothed density field on a 128$^3$ grid. Then they 
extracted 10 cubic regions, so that each one was used as initial state for their zoom-in simulations. In this paper, we 
report the average physical properties obtained considering all the cubic elements that cover a cosmological 
hydrodynamical simulation.  

Second, that the value of the ratio of kinetic energy to gravitational energy has proved to be very important 
in numerical simulations for modeling the gravitational collapse of gas cores, because it measures the relative 
importance of the kinetic energy provided initially, see for instance \cite{miyama}, \cite{hachisu1}, 
\cite{hachisu2}, and \cite{tsuribe1}. In the best case, the initial conditions for 
the studies of gas core collapse are motivated from observations, see for 
instance, ~\cite{bergin} (and references therein) which reported the physical
properties of cloud cores. Particularly, the virial parameter, which is directly related to the ratio of kinetic 
energy to gravitational energy, has recently been measured by \cite{kauffmann} who recently compiled a 
catalog of 1325 molecular gas clouds. Observational values 
of the ratio $\frac{E_{\rm kin}}{E_{\rm grav}}$ for prestellar cores
have been found to be within the range 10$^{-4}$ to 0.07, see \cite{caselli} and
\cite{jijina}. As far as we are aware, no observational estimates of this ratio for large gas 
structures, such as the ones considered in this paper, have been reported elsewhere. It can also be expected 
that the value of this ratio may be relevant in the collapse of clouds and clusters 
of clouds, see \cite{arreaga2016} and \cite{arreagaASS2017}.  
    
Third, that the usefulness of the calculations on the statistical properties of groups of gas clumps mentioned above, 
can be illustrated by the papers of \cite{perez1} and \cite{perez2}. These authors constructed 2D and 3D catalogs 
of galaxy pairs from cosmological hydrodynamical simulations, so that the 3D catalog contained 88 galaxies 
in pairs, whose statistical properties were compared with galaxies in pairs found in the 2dFGRS catalog. In 
addition, in the paper \cite{perez1} 
(\cite{perez2}) the authors investigated tidal interactions and their effects on star 
formation (on colours and chemical abundances). It can be seen that these studies can be extended 
to galaxy groups with a higher number of members, so that a statistical analysis of higher multiplicity galaxy 
groups, as the one proposed in this paper, can be useful.     
   
The rest of this paper is structured as follows. In Section \ref{sec:sim} we describe the simulation and some 
computational issues are presented in Sections \ref{subs:res} and \ref{subs:code}. The code that will be applied in 
this paper is presented in Section \ref{subsec:parti}. To characterize the code, some plots are   
described in Section \ref{subsec:parti}, which are presented in terms of the number of cubic cells. The results in 
terms on physically meaningful quantities are described in Sections \ref{sec:results}, which include the 
calculation of the mass function, the size function, and the multiplicity function of the gas clumps of the 
simulation\footnote{The term multiplicity must be understood in this paper as the the number of members 
collected into a group. The term gas clump here indicates a large 
collection of gas, which will likely form a denser structure by means of the gravitational collapse.}. In 
Section \ref{subs:betahalos} we present 
the determination of the dimensionless ratio of the 
gravitational energy to the potential energy of the gas clumps. The statistical properties of gas clump groups 
is described in Section \ref{subs:grupos}. The Section \ref{sec:discu} presents a discussion of the results 
obtained and Section \ref{sec:comp} makes a comparison with other papers to highlight the consistency of the 
results obtained and also their shortcomings. Finally, in Section \ref{sec:conclu} some 
concluding remarks are presented.  
 %%%%%%%%%%%%%%%%%%%%%%%%%%%%%%%%%%%%%%%%%%%%%%%%%%%%%%%%%%%%
\section{The simulation}
\label{sec:sim}

We consider a small part of the observable Universe, which is delimited by a cubic 
box, whose side length is $L=106$\ Mpc. The initial content
of matter is characterized by $\Omega_m=$ 0.2726 and the content of dark energy is
$\Omega_{\Lambda}=$0.7274. The sum of these quantities  $\Omega_m+ \Omega_{\Lambda}=$ 1.0, corresponding 
to a flat model of the Universe, expanding with a Hubble
parameter H$_0=100 \, h$ km s$^{-1}$ Mpc$^{-1}$.  $h$ is an indetermination factor,
given by $h=0.704$. These values that we have chosen for the content of matter and the expansion rate 
H$_0$ are the most accurate that have been determined using the most recent
observations, see \cite{planck2014} and \cite{illustris}. The average mass density in this region of
the Universe to be simulated is given by $\rho_0=$ 2.92 $\times$ 10$^{-30}$ g cm$^{-3}$ and the
initial and final redshifts are fixed at $z=127$ and $z=0$, respectively.

To generate a set of density perturbations consistent with
this cosmological model, we used the publicly available code provided by \cite{n-genic}. The simulation particles were 
initially placed in the center of 1024 partition elements of a uniform
mesh, so that an initial power spectrum $P(k)$ can be constructed by moving the simulation particles according
to the linear spectrum defined by \cite{eisenstein}, and the expected minimum and maximum
wave numbers are $k_{\rm min}=1.0 \times 10^{-6}$  and $k_{\rm max}=100 \, h$\ Mpc$^{-1}$, respectively. The
normalization of the power spectrum was fixed at a value of $\sigma_8$=0.809.

It should be noted that the code \cite{n-genic} generates the initial set of particles in pairs; that
is, for each dark-matter particle there is a
gas particle. So that the number $N_{\rm DM}$ of dark-matter (DM) particles and the 
number $N_{\rm G}$ of gas (G) particles are both equal to $23,887,872$.
Hence, the particles have masses given by $m_{DM}= 3.18 \times 10^{9} M_{\odot}$ and
$m_{G}= 6.4 \times 10^{8} M_{\odot}$, respectively. The time evolution of the simulation up to $z=0$
required a little more than 5,000 CPU hours, running 
in 250 processors in the cluster Intel Xeon E5-2680 v3 at 2.5 Ghz of
LNS-BUAP. The computational method to be described in \ref{subsec:parti} and whose results to be  
described in \ref{sec:results}, will be applied only to last output, that is, the snapshot at redshift $z=0$.   

%%%%%%%%%%%%%%%%%%%%%%%%%%%%%%%%%%%%%%%%%%%%%%%%%%%%%%%%%%%%%%%%%%%%%%%%%%%%%%%%%%%%%%%%%%%%%%%%%%%%%%%%%%
\subsection{Resolution and equation of state}
\label{subs:res}

The resolution of a simulation can be characterized by the Jeans
wavelength $\lambda_J=\sqrt{ \frac{\pi \, c^2}{G\, \rho}}$ where $c$ is the instantaneous speed of 
sound and $\rho$ is the local density and $G$ is Newton gravitational constant. A more useful form 
for a particle-based code, is the Jeans mass, $M_J$, which is given 
by $M_J \equiv \frac{4}{3}\pi \; \rho \left(\frac{ \lambda_J}{2}
\right)^3 = \frac{ \pi^\frac{5}{2} }{6} \frac{c^3}{ \sqrt{G^3 \,
\rho} }$.

The values of the density and speed of sound must be updated according to
the ideal equation of state $p= c^2 \, \rho$.The average gas temperature at $z=127$ is $T=245$ K, so 
the ideal speed of sound can be obtained by the relation $c=\sqrt{\gamma \, k_B \, T/m_p}$ 
where $\gamma\, \equiv 5/3$, $k_B$ is the Boltzmann constant and $m_p$ is the proton mass. From these relations we 
obtain $\lambda_J=0.16 \, h^{-1} $ Mpc and the Jeans mass $M_J= 3 \times 10^{8}\ M_{\odot}$. The smallest 
mass particle that a SPH calculation must resolve to be reliable is given 
by $m_r \approx M_J / (2 N_{\rm neigh})$, where $N_{\rm neigh}$ is the number of neighboring 
particles included in the SPH kernel; see \cite{bateburkert97}.The
ratio of the simulation mass $m_G$ calculated above and this resolution mass $m_r$ is then given 
by $m_G/m_r= 569$ for the simulation.
%%%%%%%%%%%%%%%%%%%%%%%%%%%%%%%%%%%%%%%%%%%%%%%%%%%%%%%%%%%%%%%%%%%%%%%%%%%%%%%%%%%%%%%%%%%%%%%%%
\subsection{The evolution code}
\label{subs:code}

The simulations of this paper are generated by the particle-based code Gadget2, which is
based on the SPH method for evolving the particles according with the Euler equations of
hydrodynamics; see~\cite{springel}. The Gadget2 has a Monaghan-Balsara form 
for the artificial viscosity, see \cite{balsara1995}, so that the strength of the
viscosity is regulated by setting the parameter $\alpha_{\nu} = 0.75$ and
$\beta_{\nu}=\frac{1}{2}.
\times \alpha_v$, see Equations 11 and 14
in~\cite{springel}. The Courant factor has been fixed at $0.1$.

The SPH sums are evaluated using the spherically symmetric M4 kernel of \cite{monalatt}, 
and so gravity is spline-softened with this same kernel. 
The smoothing length $h$ establishes the compact support so 
that only a finite number of neighbors to each 
particle contribute to the SPH sums. The smoothing length changes with time for each particle 
so that the mass contained in the kernel volume is a constant for the estimated density. Particles also 
have gravity softening lengths $\epsilon$, which change step by step  
with the smoothing length $h$, so that the ratio $\epsilon/h$ is of
order unity. In Gadget2, $\epsilon$ is set equal to the minimum 
smoothing length $h_{\rm min}$, calculated over all particles at the end of each time
step. It must be noted that the upper bound of the softening length used in Gadget2 code to 
run the simulations of this manuscript is 0.02 Mpc.

%%%%%%%%%%%%%%%%%%%%%%%%%%%%%%%%%%%%%%%%%%%%%%%%%%%%%%%%%%%%%%%%%%%%%%%%%%%%%%%%%%%%%
\subsection{The cubic partitions}
\label{subsec:parti}

Our code makes a partition of the simulation box by means of a set of identical cubic 
elements. Let us define the level $l$ of 
the partition, so that the number of partition elements per side of the 
simulation box is given by 2$^l$. In this paper, only partition levels 6 and 7 will be 
considered, so that the number of length elements per simulation side are 64 and 
128, respectively. For these partitions, the total number of identical cubic elements in 
which the entire simulation volume is divided are therefore ($2^l)^3 \, \equiv $ 262144 
and 2097152, respectively. Let us label these partitions as P6 and P7, respectively. It should 
be noted that the side length of each cubic element of the partitions P6 and P7 is 1.66 
Mpc and 0.83 Mpc, respectively.  

We next determine the average gas density inside each cubic element of the partitions, by using 
a method often found in particle simulations, which is the nearest grid point (NGP) method, see \cite{charles}. It 
deposits the entire mass of the particle to the nearest grid point. For the partitions 
P6 and P7, the number of cubic elements with a non-negligible number of particles is around 130,000 and 
1,000,000, respectively. Let us define the 
normalized density of the cubic element by ${\rm lrho0}=\log_{10}\left(\rho/\rho_0 \right)$, where $\rho_0$ is 
the average mass density of the Universe, as described in Section \ref{sec:sim}. Then, the average density of 
this set of cubic elements is $<{\rm lrho0}>=0.23$ and $<{\rm lrho0}>=0.026$, respectively. The 
standard deviation of the normalized density is $\sigma_{\rm lrho0}=0.89$ and 
$\sigma_{\rm lrho0}=1.05$, respectively.   

\begin{figure}
\begin{center}
\includegraphics[width=4.0 in]{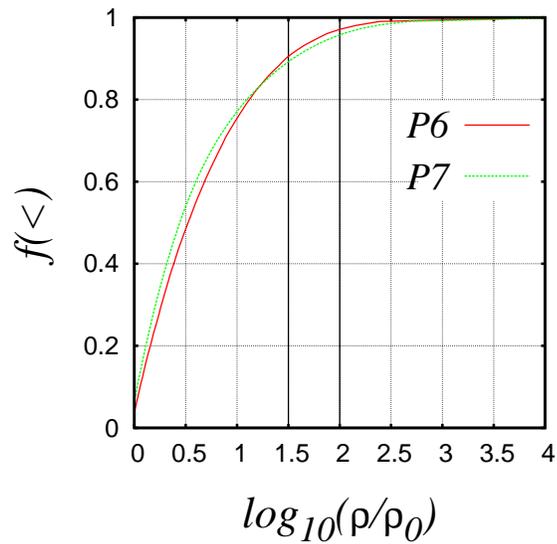}
\caption{\label{NpDen} The fraction of the number of cubic elements having a ratio of the normalized density with a value 
less than the value of the horizontal axis. The vertical lines at 1.5 and 2 determine the density threshold of the 
chosen cubic elements, as defined in the text, see Section \ref{subsec:parti}.}
\end{center}
\end{figure}

To start our study, we arbitrarily chose the initial values for the minimum normalized density, 
denoted by ${\rm lrho0}_{\rm min}$, so that in this paper, only the two values 
of ${\rm lrho0}_{\rm min}$=1.5 and 2 will be considered. We will focus only on those cubic elements of the 
partition whose normalized density is greater than 
this value of ${\rm lrho0}_{\rm min}$. In Fig.\ref{NpDen}, the values 
of ${\rm lrho0}_{\rm min}$ are shown in the horizontal axis, so that all the cubic elements of the 
partitions located to the right-hand side of those vertical lines will be defined as 
the {\it chosen cubic elements}; that is, they 
satisfy the condition $\log_{10}\left(\rho/\rho_0 \right) > {\rm lrho0}_{\rm min}$. These values of 
${\rm lrho0}_{\rm min}$=1.5 and 2 correspond to an overdensity $\Delta$ of 31 and 316 times the average 
matter density of the Universe, $\rho=\Delta \, \rho_0$. It must be emphasized that this overdensity 
value $\Delta$ is determined such that the matter with a density $\rho$ greater than $\Delta \, \rho_0$ is 
virialized in a given cosmology. The most commonly used value in simulations for the overdensity is $\Delta=$200, 
see \cite{halosmad}.

%%%%%%%%%%%%%%%%%
%%%%%%%%%%%%%%%      

Some of the properties of these sets of chosen cubic elements of 
the partitions are shown in Table~\ref{tab:stat}, as follows: column one shows the label of each 
partition; column two shows the partition level defined above and the number 
of identical cubic elements; column three shows the minimum density value used to define 
the set of chosen cubic elements; column four shows the number of chosen cubic elements per 
each partition; column five shows the number of cubic elements that are linked as neighbors, see 
Section~\ref{subs:funcionmult} below and column six shows the peak number of gas particles 
found in only one cubic element of this set of chosen cubic elements.        

\begin{figure}
\begin{center}
\includegraphics[width=4.0 in]{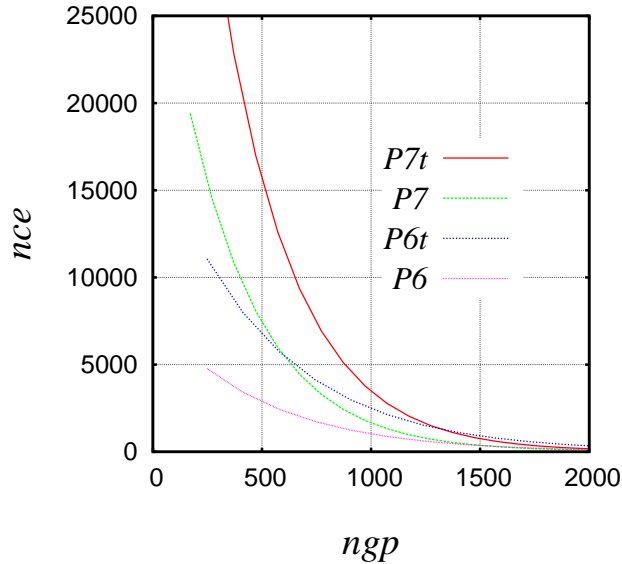}
\caption{\label{fig:FuncDistNpE} The distribution function of the number of gas particles 
(shown in the horizontal axis) inside each chosen cubic element (shown in the vertical axis).}
\end{center}
\end{figure}

It should be mentioned that the number of particles inside each cubic element can also be used as the selection 
criterion to define the chosen cubic elements. In this case, there would be a 
minimum particle number given before-hand, so that only those cubic elements with a greater number of particles would be 
considered in the calculation of properties. However, the results are more interesting when the criterion is based on 
density. In Fig.~\ref{fig:FuncDistNpE} we show the distribution function of the number of cubic elements against 
the number of gas particles inside each cubic element.

%%%%%%%%%%%%%%%
\begin{figure}
\begin{center}
\includegraphics[width=3.0 in, height=3.0 in]{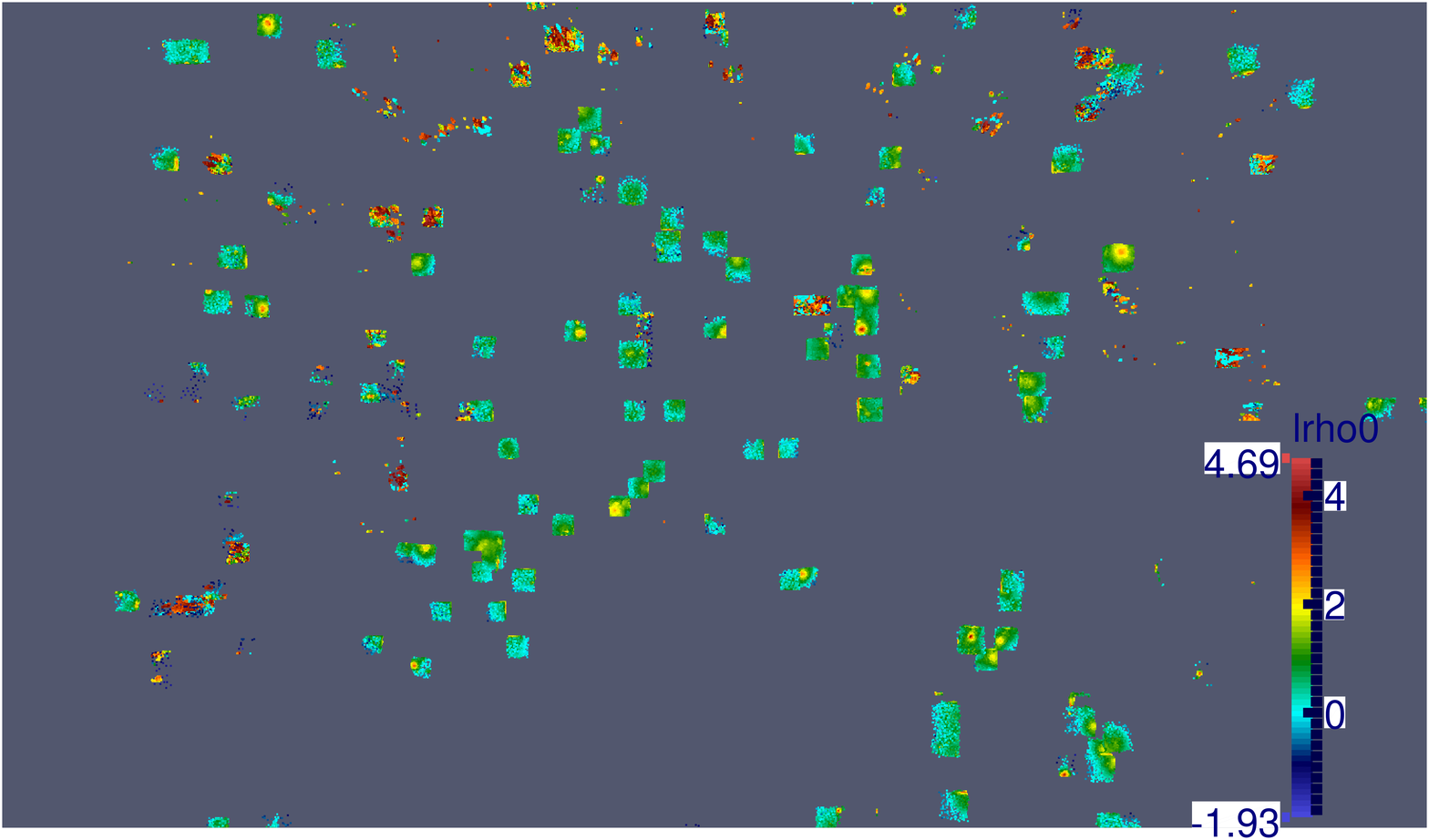}
\caption{\label{EscoT_006} Spatial distribution of all the cubic elements satisfying the density condition 
defined in Section \ref{subsec:parti} using partition P6. The length of the side is 100 Mpc. }
\end{center}
\end{figure}

\begin{figure}
\begin{center}
\includegraphics[width=3.0 in, height=3.0 in]{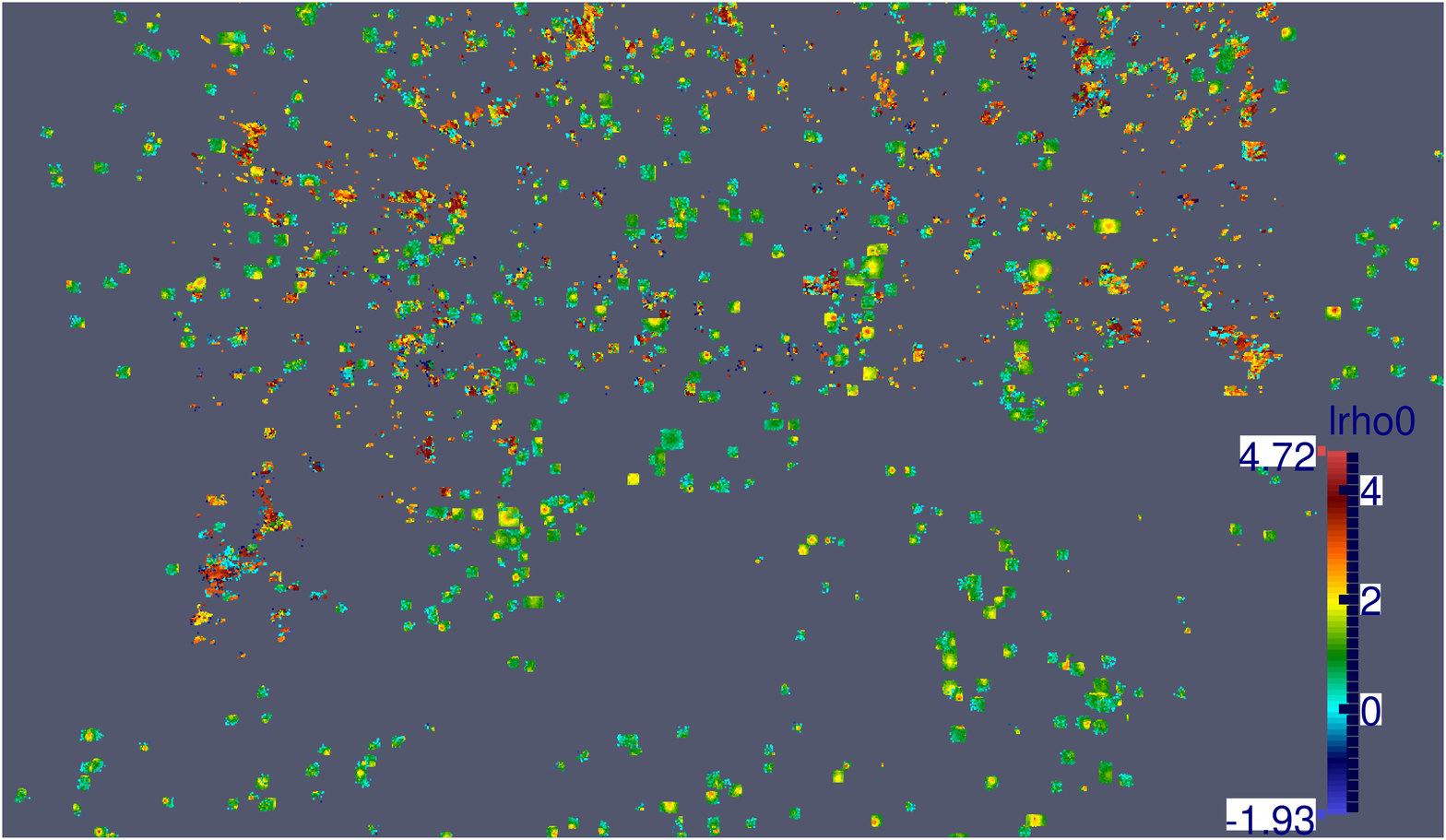}
\caption{\label{EscoT_007} Spatial distribution of all the cubic elements satisfying the density condition 
defined in Section \ref{subsec:parti} using the partition P7. The length of the side is 100 Mpc.}
\end{center}
\end{figure}

%%%%%%%%%%%%%%%
The spatial distribution of the identical cubic elements defined in this section is 
shown Figs. \ref{EscoT_006} and \ref{EscoT_007}, in which the normalized density value appears 
in the right-hand column of each plot as ${\rm lrho0}$.

%%%%%%%%%%%%%%%%
%%%%%%%%%%%%%%%%%
%%%%%%%%%%%%%%%%%%%%%%%%%%%%%%%%%%%%%%%%%%%%%%%%%%%%%%%%%%%%%%%%%%%%%%%%%%%%%%%%%%%%%%%
\section{Results}
\label{sec:results}

To highlight the spatial scale of the simulation and the code presented in this 
paper, let us now summarize, very briefly, the well-established 
current scenario about the formation of structure in cosmological simulations. 
At the scale of 100 Mpc and redshift z=0, our simulations will produce a cosmic network of dense
filaments. At a scale of 
10 Mpc, the most massive dark-matter halos will be formed mainly at the 
intersection of those filaments. At a scale of a few Mpc, the gas will condense
and start forming virialized regions in the center of the
collapsed dark-matter halos, so that galaxy clusters will appear. At a
scale of a few kpc, the dense gas will form galaxies, so that at the
nucleus of some of the galaxies, at the scale of a few pc, the gas
will collapse gravitationally to start forming stars. 

In Sections \ref{subs:funcionmasa}-\ref{subs:funcionmult}, we will characterize the set of chosen cubic 
elements defined in Section \ref{subsec:parti}. The gas particles 
contained in these chosen cubic elements of the partitions will 
be considered as an approximation of the distribution of the gas clumps, as is explained in 
Sections \ref{subs:grupos} and \ref{subs:betahalos}. It will be seen that 
an approximate size of these gas clumps is of the order of 1 Mpc 
or less, while the groups formed by these gas clumps are of the order of a few Mpc, and can therefore 
be identified with the structures forming virialized regions of gas (as mentioned in the previous paragraph).        

%%%%%%%%%%%%%%%%%%%%%%%%%%%%%%%%%%%%%%%%%%%%%%%%%%%%%%%%%%%%%%%%%%%%%%%%%%%%%%%%%%%%%%%%
\subsection{The distribution function of the mass for the chosen cubic elements}
\label{subs:funcionmasa}

\begin{figure}
\begin{center}
\begin{tabular}{c}
\includegraphics[width=3.5 in]{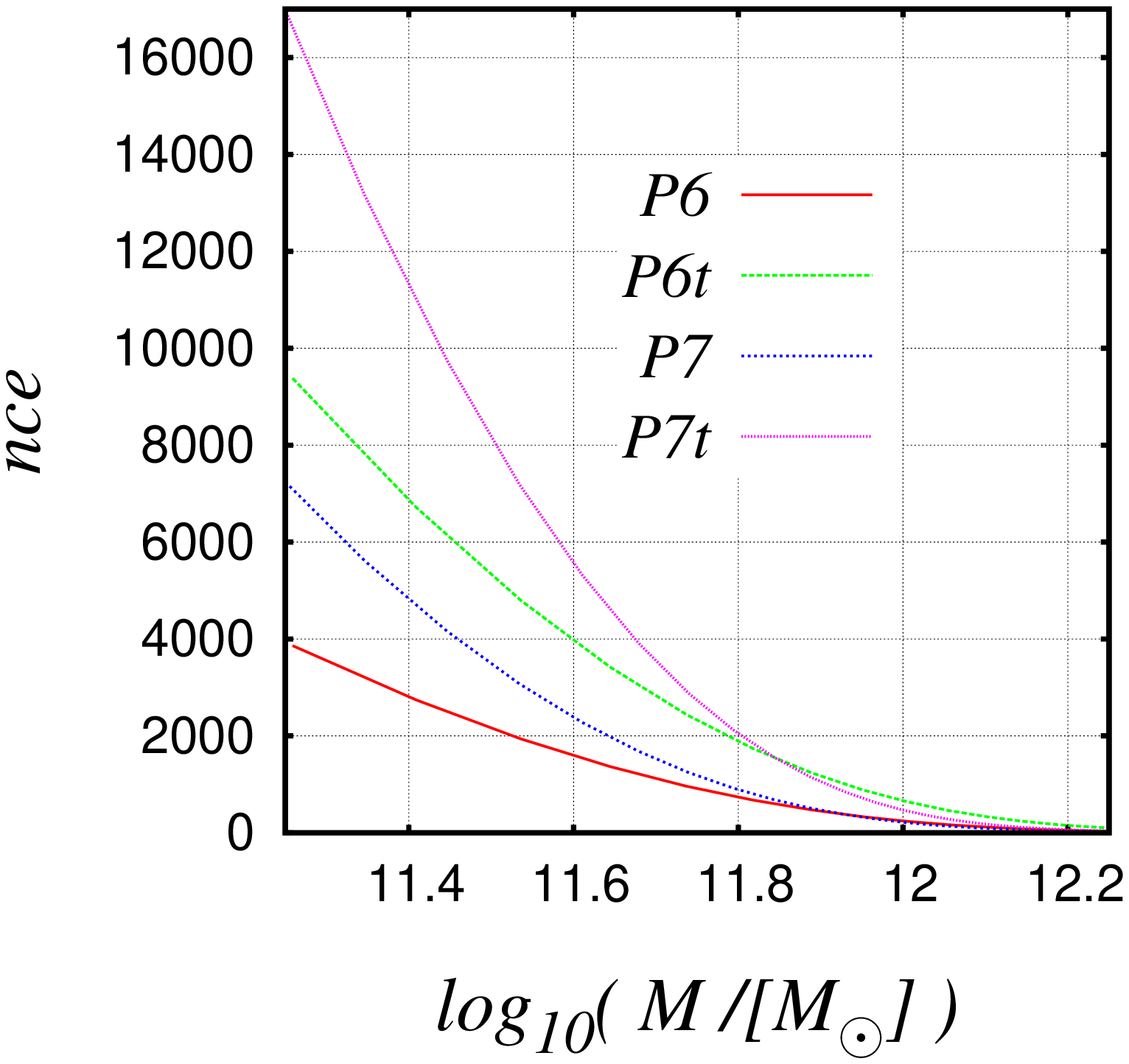} \vspace{0.1 cm} \\
\includegraphics[width=3.5 in]{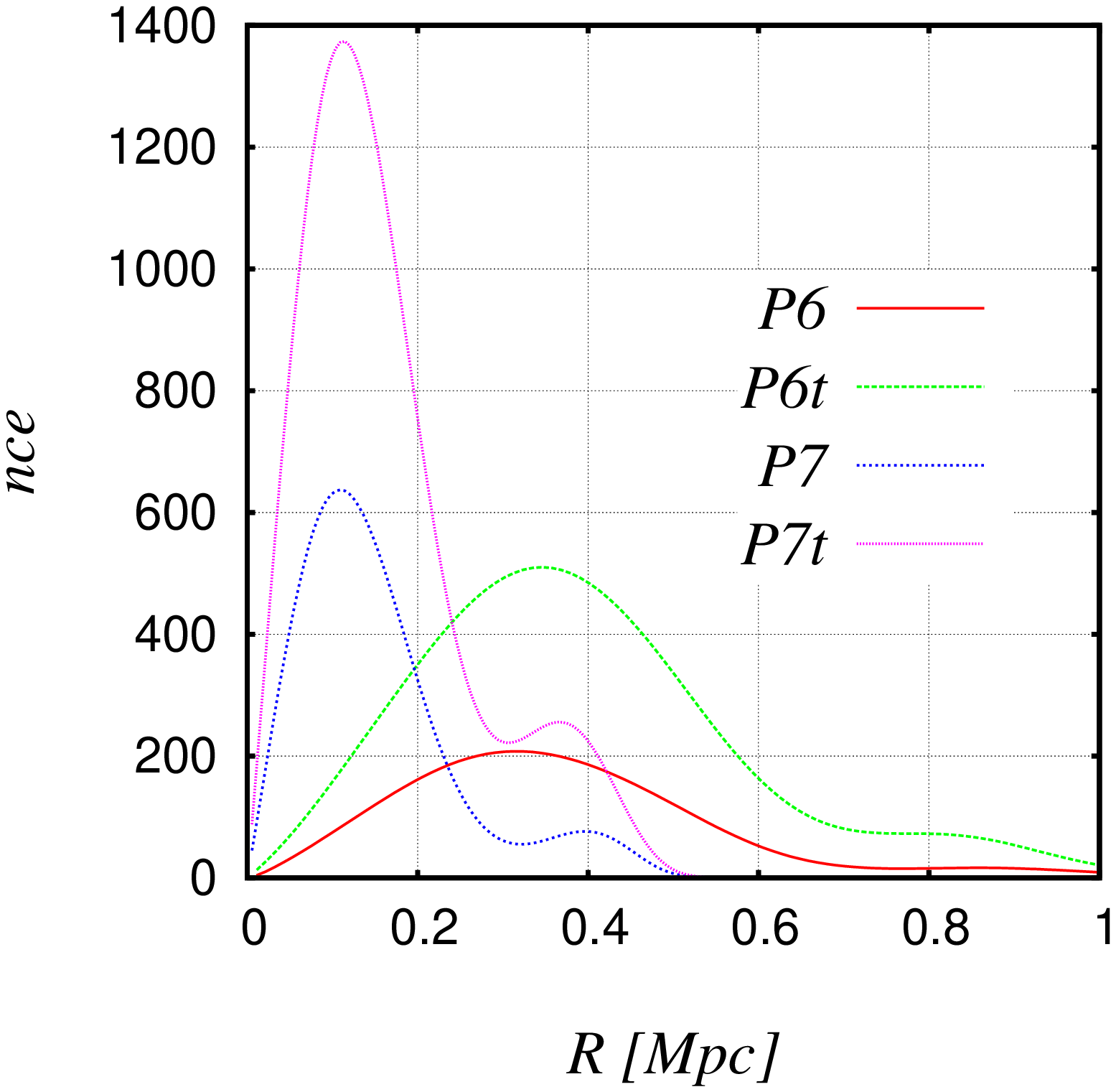} \vspace{0.1 cm}
\end{tabular}
\caption{\label{Fig:IMRF} (top) The mass distribution function for the chosen cubic elements of the partitions. The vertical 
axis shows the number of chosen cubic elements having a mass shown in the horizontal axis. (bottom) The 
size distribution function for the chosen cubic elements of the partitions. The vertical axis shows the 
number of chosen cubic elements having a radius shown in the horizontal axis.}
\end{center}
\end{figure}

We first count the gas particles within each chosen cubic element of the partition and thus immediately have 
the mass contained. We determine the minimum and maximum masses of this set of gas particles and make 
a mass partition in terms of $n_{\rm bin}=50$ bins, so that we count 
all the chosen cubic elements with a mass within each mass bin. The resulting distribution 
function of the mass of the chosen cubic elements is shown in the 
top panel of Fig.\ref{Fig:IMRF}.

%%%%%%%%%%%%%%%%%
%%%%%%%%%%%%%%%%%

The behavior of these curves are all similar, as expected: they are quite separated for the 
smallest mass scale, indicating that there is a large difference in the number of 
low-mass chosen elements found in each partition. The smallest mass scale 
identified for all the partitions is around $\log_{10} \left( M/M_{\odot} \right)\, \approx 11.2$, whereas 
the largest mass scale, according to column six of Table~\ref{tab:stat} is around 13, although the 
number of chosen cubic elements decreases significantly in the plot around a mass scale of 12.2.

It appears that the pair of curves for partitions P6-P7 and P6t-P7t are closer to each other. This indicates 
that for the mass determination, the resolution parameter of the partitions seems to be more important 
than the density threshold parameter.  

%%%%%%%%%%%%%%%%%%%%%%%%%%%%%%%%%%%%%%%%%%%%%%%%%%%%%%%%%%%%%%%%%%%%%%%%%%%%%%%%%%%%
\subsection{The distribution function of the radius for the chosen cubic elements}
\label{subs:funcionradio}

Let us now estimate a distribution function of the size for the dense gas contained 
in each chosen cubic element. We again re-consider the gas particles determined 
in Section \ref{subs:funcionmasa} to calculate the center of mass and 
then locate the particle that is furthest from this center of mass but still inside 
the chosen cubic element, so that half of this distance is defined as a geometrical measure of 
the set of gas particles contained in the cubic element. We next make a radial partition of 
the radii thus obtained in terms of $n_{\rm bin}=50$ radial bins, so that a distribution function 
can be obtained by this procedure, as shown in the bottom panel 
of Fig.\ref{Fig:IMRF}. 

Let us emphasize that the radius of the horizontal axis in both panels of Fig. \ref{Fig:IMRF} is 
just a label to identify the shells of the radial partition. Let $rmax$ be the maximum distance of the furthest 
particle from the center of mass, as mentioned in the paragraph above. Then, for the 
grids P7 and P7t we have $rmax_{P7}$ and for the grids P6 and P6t we have $rmax_{P6}$. As expected, we always have 
$rmax_{P7} < rmax_{P6}$. This indicates that the grids P6 and P6t have radial shells with a width larger than the 
radial shells of the grids P7 and P7t. For this reason, the curves for the grids P6 and P6t are 
located to the right side of the curves for the grids P7 and P7t. 

It should be noted that the partitions P6 have, in general, a core radius that is larger than those of the partitions 
P7, so that the average radius is around 0.3 and 0.1 Mpc, respectively. The fixed size of the cubic 
element of each partition, as explained in Section \ref{subsec:parti}, 
determines an upper limit of the distribution function of the radius. However, the radii 
detected by this procedure are quite smaller than this 
upper limit. 

It appears that the pair of curves for the partitions P6-P6t and P7-P7t are closer to each other. This 
indicates that for the size determination, the density threshold parameter of the partitions seems to be more important 
than the resolution parameter.    

%%%%%%%%%%%%%%%%%%%%%%%%%%%%%%%%%%%%%%%%%%%%%%%%%%%%%%%%%%%%%%%%%%%%%%%%%%%%%%%%%%%%%
\subsection{The distribution function of the multiplicity for the chosen cubic elements}
\label{subs:funcionmult}

To take further advantage of the partitions described in Section \ref{subsec:parti}, we now 
determine the chosen cubic elements located in the same neighborhood. To do this, we 
first define a proximity parameter, which must be a distance. We start this calculation 
by arbitrarily using the hypotenuse of the cubic element of the partitions, as defined in 
Section \ref{subsec:parti}, so that in the case of the 
partitions P6, this proximity parameter is fixed at 1.44; for the partitions P7 
it is fixed at 2.88 Mpc. 

\begin{figure}
\begin{center}
\includegraphics[width=3.5 in]{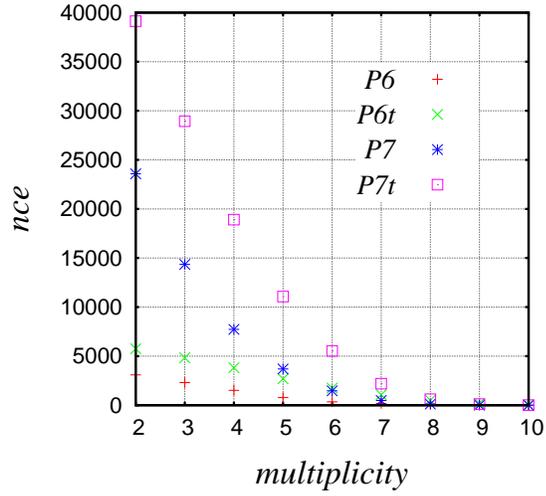}
\caption{\label{fig:Mult} The multiplicity distribution function of the number of chosen cubic elements (shown in the vertical 
axis) having the number of neighbors (shown in the horizontal axis).}
\end{center}
\end{figure}

We next determine all of the chosen cubic elements whose 
geometric centers are separated by a distance smaller than or equal to this proximity 
parameter. Let us define the multiplicity as the total number of cubic elements that are all 
linked to each other in this way, the 
result of this procedure is shown in Fig.\ref{fig:Mult}. Following this 
procedure, we have a list of neighbors for each chosen cubic element, which will be used 
again in Section \ref{subs:grupos}.

%%%%%%%%%%%%%%%%%%
%%%%%%%%%%%%%%%%%%

\begin{figure}
\begin{center}
\includegraphics[width=4.0 in]{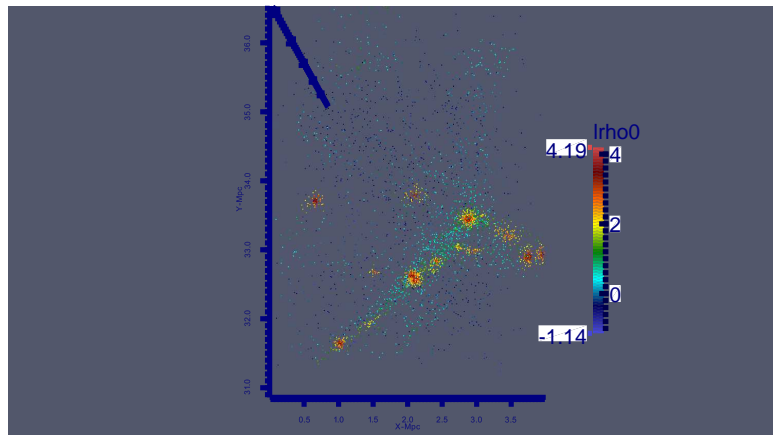}
\caption{\label{fig:ElGrupo} The group of chosen cubic elements with maximum multiplicity of 10 members, detected with  
both partitions P7 and P7t. Gas particles with a lower normalized density are allowed to 
enter in this plot to increase the total number of particles and improve the view.}
\end{center}
\end{figure}

The case of groups with multiplicity 10 is remarkable, because many of these groups have been detected 
simultaneously with the same partition. For instance, the partition P6 detects five 
groups of multiplicity 10; 38 were detected with the partition P6t; only two were detected with the P7 and 
13 were detected with the P7t. An example of such a highest multiplicity group is shown in Fig.\ref{fig:ElGrupo}, where 
it can be noted that the gas clumps follow a filament. In addition, it 
should also be noted that the group shown in Fig.\ref{fig:ElGrupo} has been detected with 
both partitions P7 and P7t.

%%%%%%%%%%%%%%%%
%%%%%%%%%%%%%%%%   

Column five of Table~\ref{tab:stat} shows the total number of 
chosen cubic elements that are grouped in the some multiplicity level detected by this 
procedure. These numbers indicate that about 74 percent of the chosen cubic elements for
partition P6 are placed into groups while 82 percent are placed in partition P6t. In the case of the 
partitions P7 and P7t, these fractions decreases to 65 percent and 74 percent, respectively.

\begin{table}[ph]
\caption{The partitions and some properties}
{\begin{tabular}{|c|c|c|c|c|c|} \hline
partition label & partition level (nct) & lrho$_{\rm min}$ & nce & nce$_{\rm Mult}$ & ngp$_{\rm max}$ \\
\hline
%                & (number of elements) &                  & (chosen cubic elements)     &           &            &          & \\
\hline
P6       & 6 (64$^3$)  &  2.0  & 6795  &  5060   &  24801   \\
\hline
P7       & 7 (128$^3$) &  2.0  & 50537 &  32891  &  17017     \\
\hline
P6t      & 6 (64$^3$)  &  1.5  & 14408 &  11874  &  24801    \\
\hline
P7t      & 7 (128$^3$) &  1.5  & 85810 &  64192  &  17107    \\
\hline
\hline
\end{tabular} }
\label{tab:stat}
\end{table}       
%%%%%%%%%%%%%%%%%%%%%%%%%%%%%%%%%%%%%%%%%%%%%%%%%%%%%%%%%%%%%%%%%%%%%%%%%%%%%%%%%%%%%%%%%%%%%%%%%%%%%%%%%%%%%%
%%%%%%%%%%%%%%%%%%%%%%%%%%%%%%%%%%%%%%%%%%%%%%%%%%%%%%%%%%%%%%%%%%%%%%%%%%%%%%%%%%%%%%%%%%%%%%%%%%%%%%%%%%%%
\subsection{Statistical properties of gas clump groups}
\label{subs:grupos}

To relate the gas particles contained in 
a chosen cubic element with a gas clump, in this section we impose two conditions on these gas particles so that they 
belong to a gas clumps if and only if: (i) the number of gas 
particles is greater than 10 and (ii) only gas particles with $\Phi_{i}<0$ will be considered to be 
in a gas clump. 

To begin with the characterization of the gas clumps and their grouping properties, we first 
re-consider the list of neighbors obtained in Section \ref{subs:funcionmult}. In addition, the proximity 
parameters for all the partitions are the same as those that were used in Section \ref{subs:funcionmult}.   

\begin{figure}
\begin{center}
\begin{tabular}{cc}
\includegraphics[width=3 in]{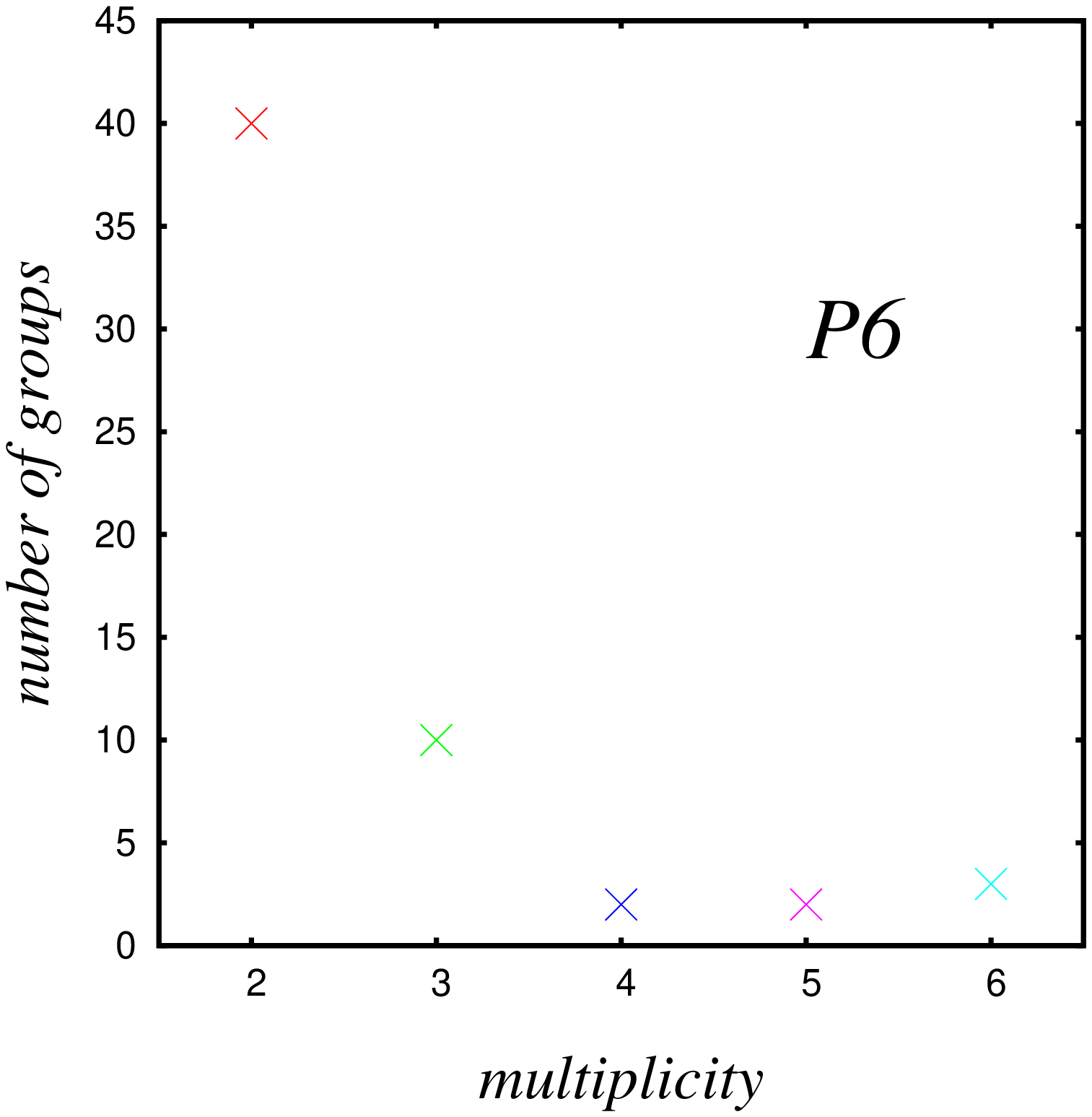} & \hspace{-1 cm} \includegraphics[width=3 in]{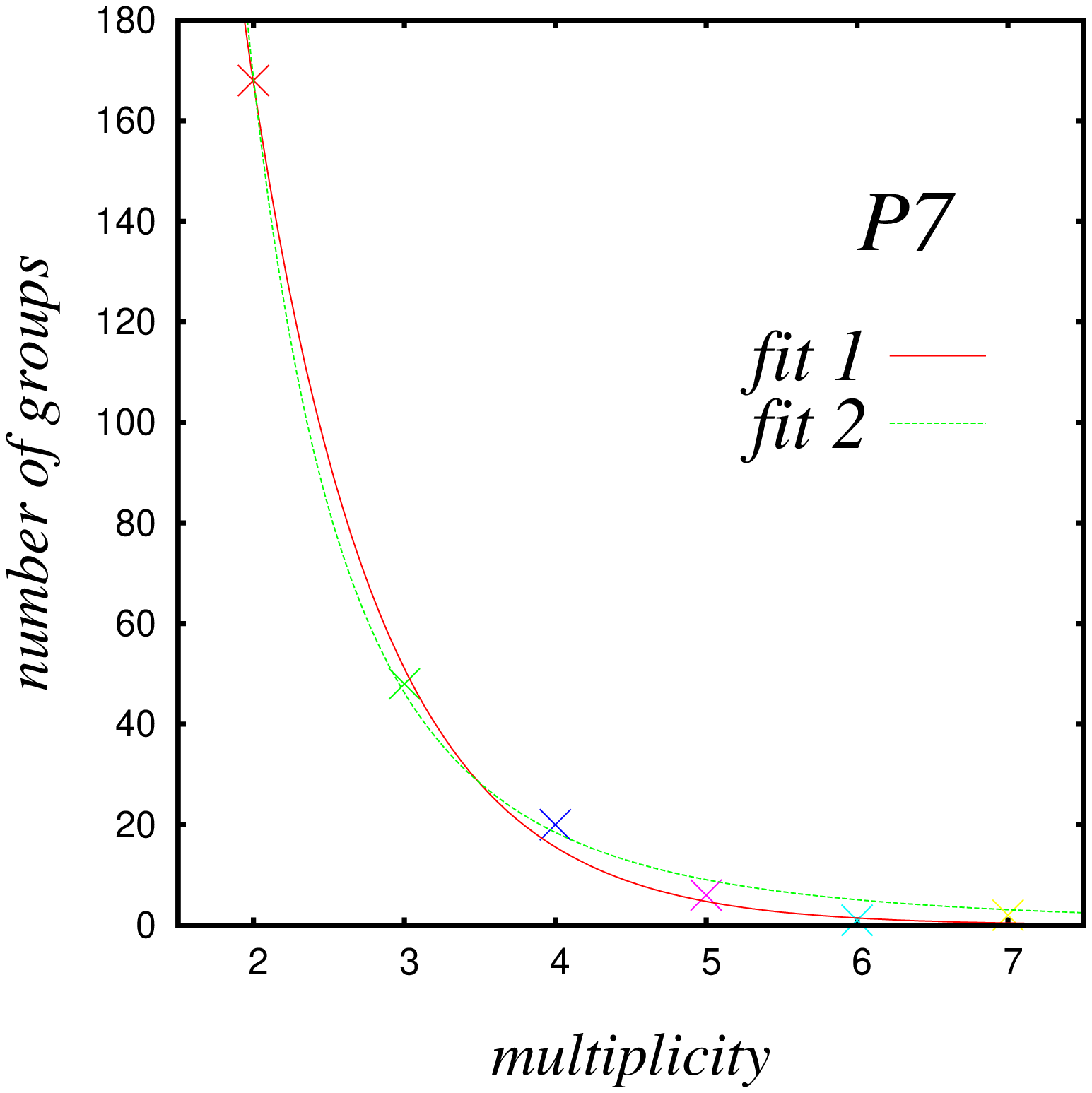}\\
\includegraphics[width=3 in]{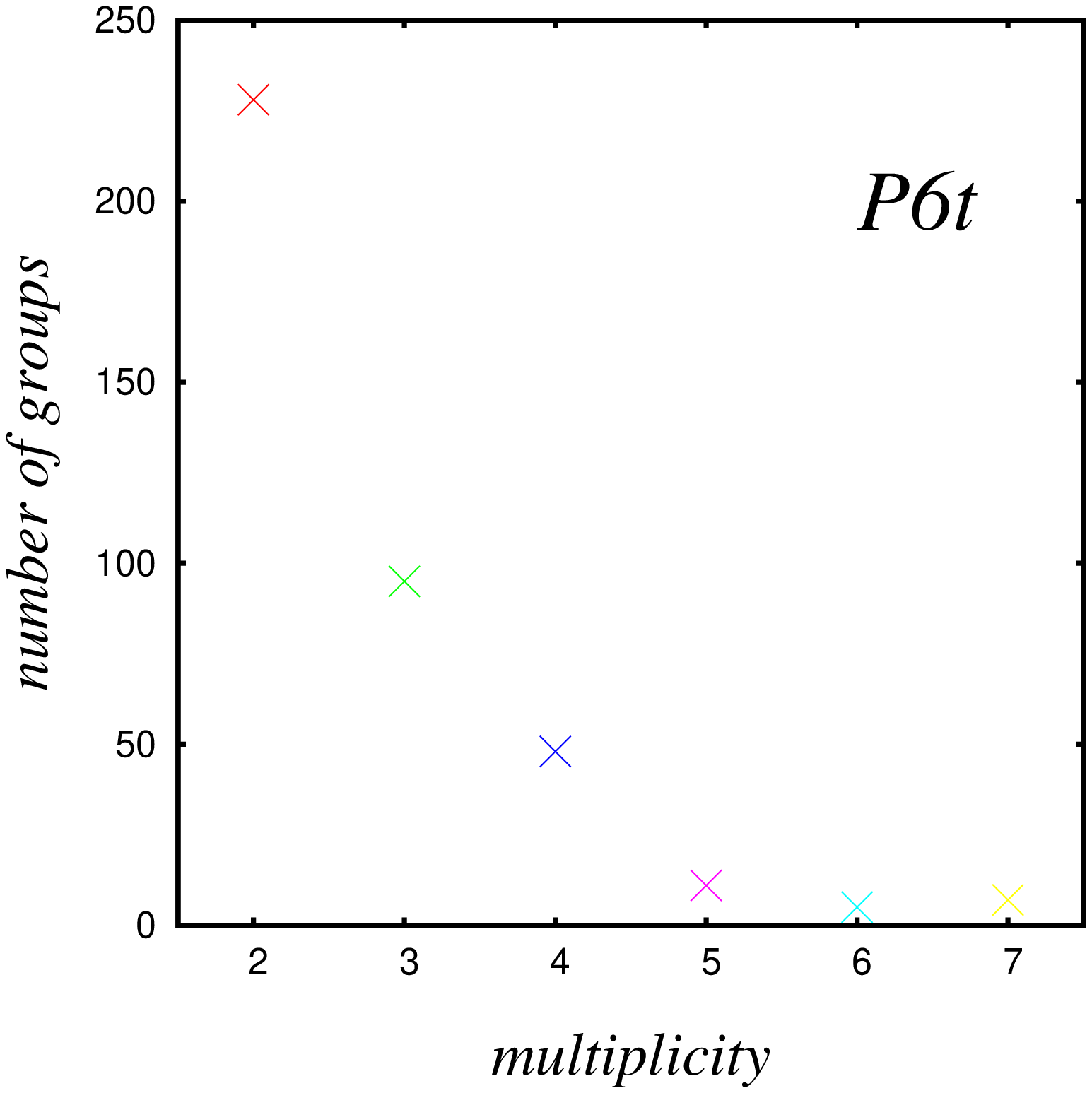} & \hspace{-1 cm} \includegraphics[width=3 in]{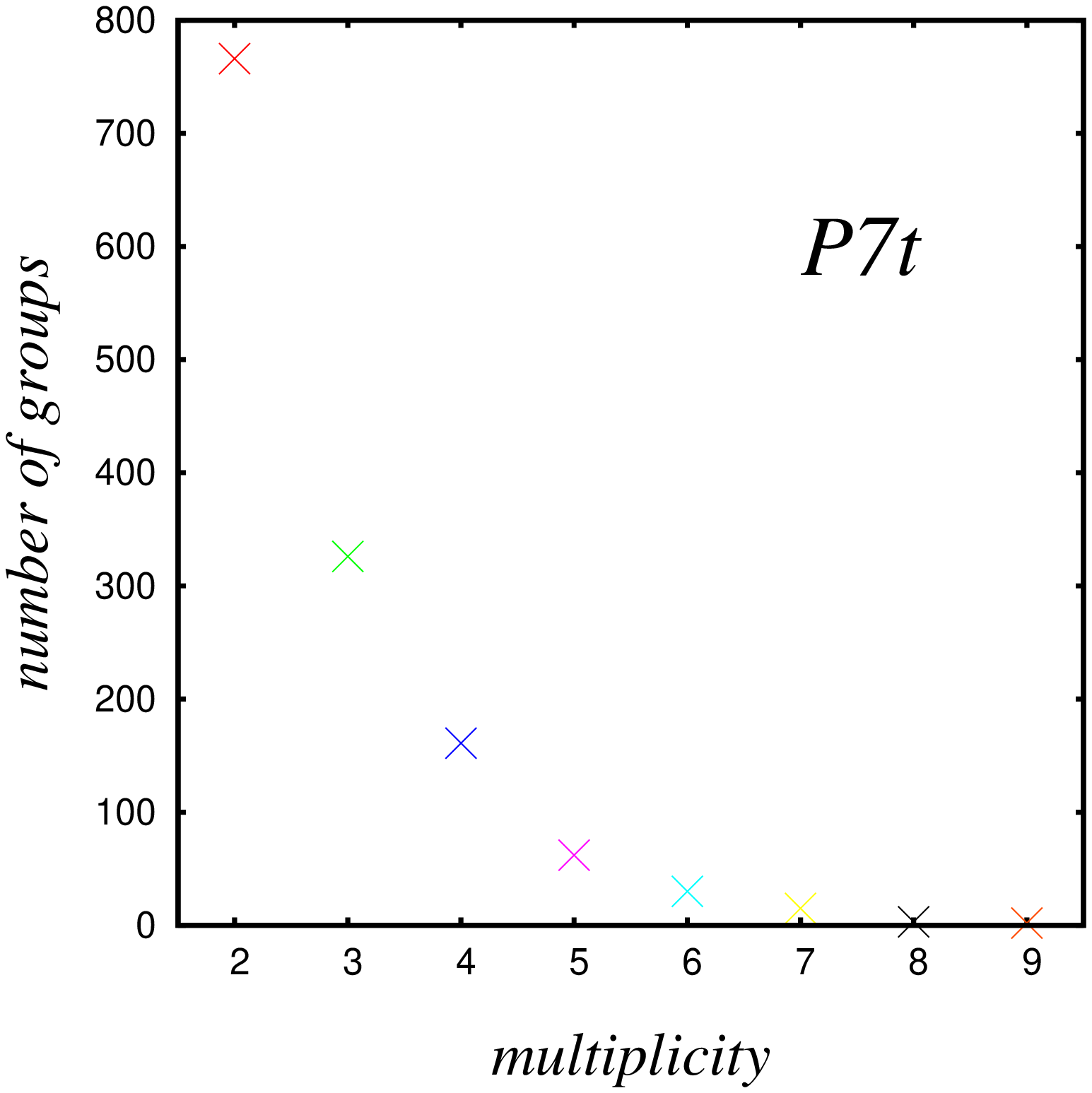}
\end{tabular}
\caption{\label{Fig:PropGru_Contf} The number of groups (ng) detected are shown in the y-axis (using the point symbol x) in terms of 
the multiplicity (m) shown in the x-axis. Two lines are included in the top-right panel to fit the data with the following 
formulae: $ng1(m) = I_1*\exp(-m/ms)$ (fit 1) and $ng2(x) = I_2*(x/ms_2)^{-n_2}$ (fit 2). The values of the fitting parameters 
are $I_1= 1802.23$, $m_s= 0.841847$, $I_2= 17.1243$, $n_2=3.18728$ and $m_{s2}= 4.09685$.}
\end{center}
\end{figure}

The application of these conditions on the list of neighbors has the immediate result that the
number of groups per multiplicity decreases significantly, as can 
be seen in Fig.\ref{Fig:PropGru_Contf}, which must be compared to 
Fig.\ref{fig:Mult}. In addition, the groups of multiplicity 10 
are no longer detected for any of the partitions. Consequently, in the x-axis of 
Fig.\ref{Fig:PropGru_Contf}, a multiplicity number is shown if and only if 
a non-zero number of groups is detected. 

%%%%%%%%%%%%%%%%%%%%%%
%%%%%%%%%%%%%%%%%%%%%% 

\begin{figure}
\begin{center}
\begin{tabular}{cc}
\includegraphics[width=3 in]{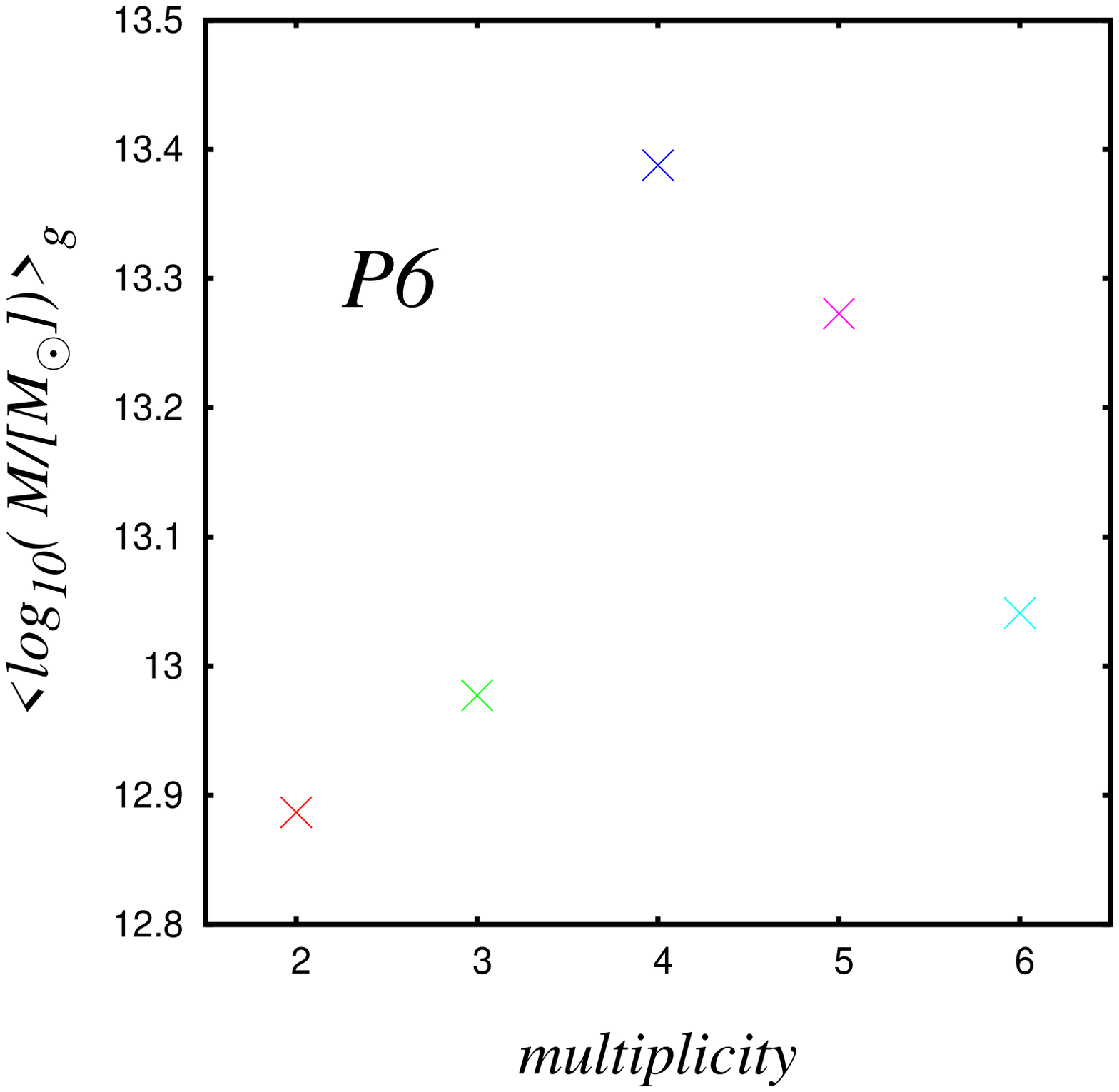} & \hspace{-1 cm} \includegraphics[width=3 in]{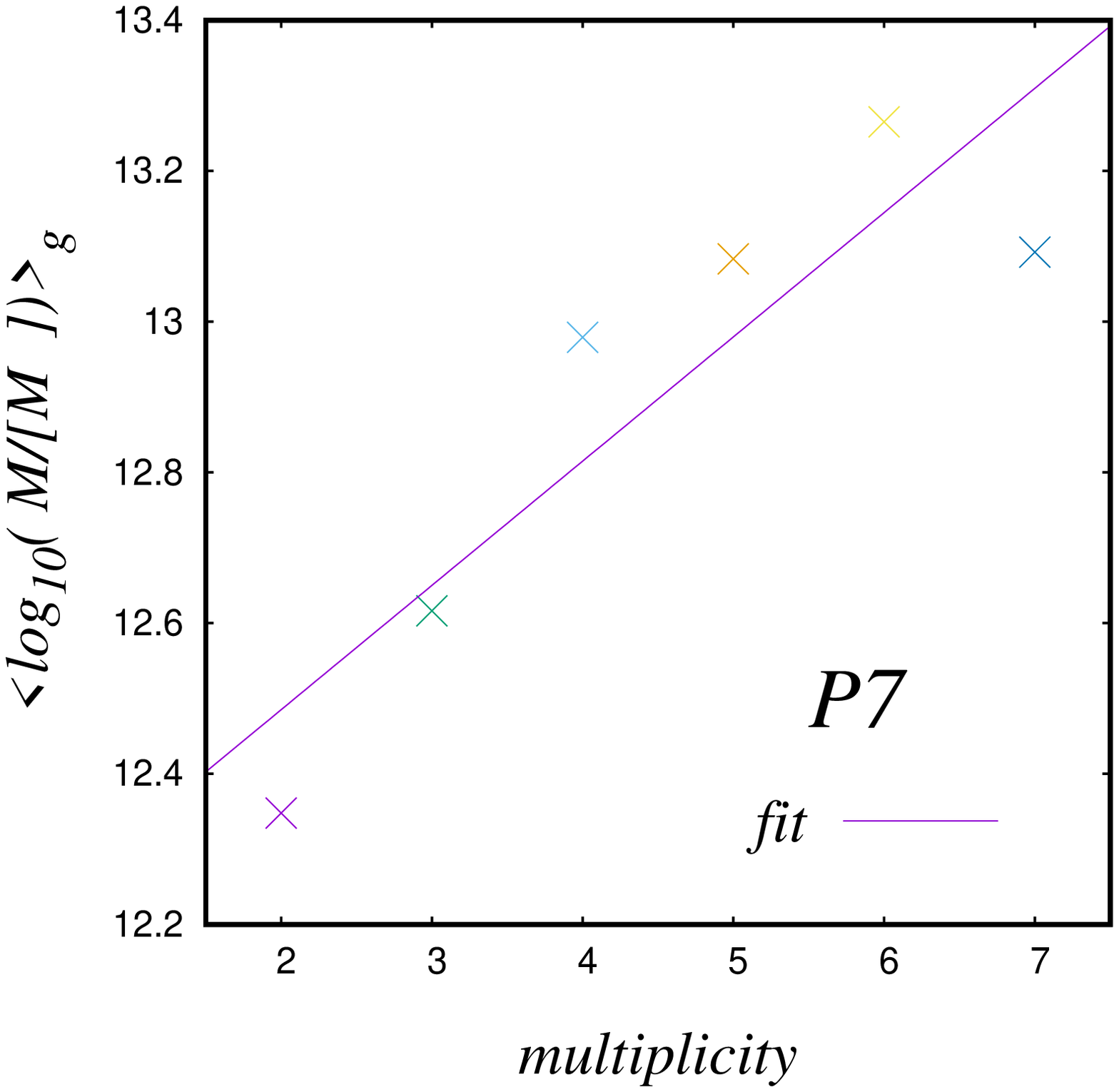}\\
\includegraphics[width=3 in]{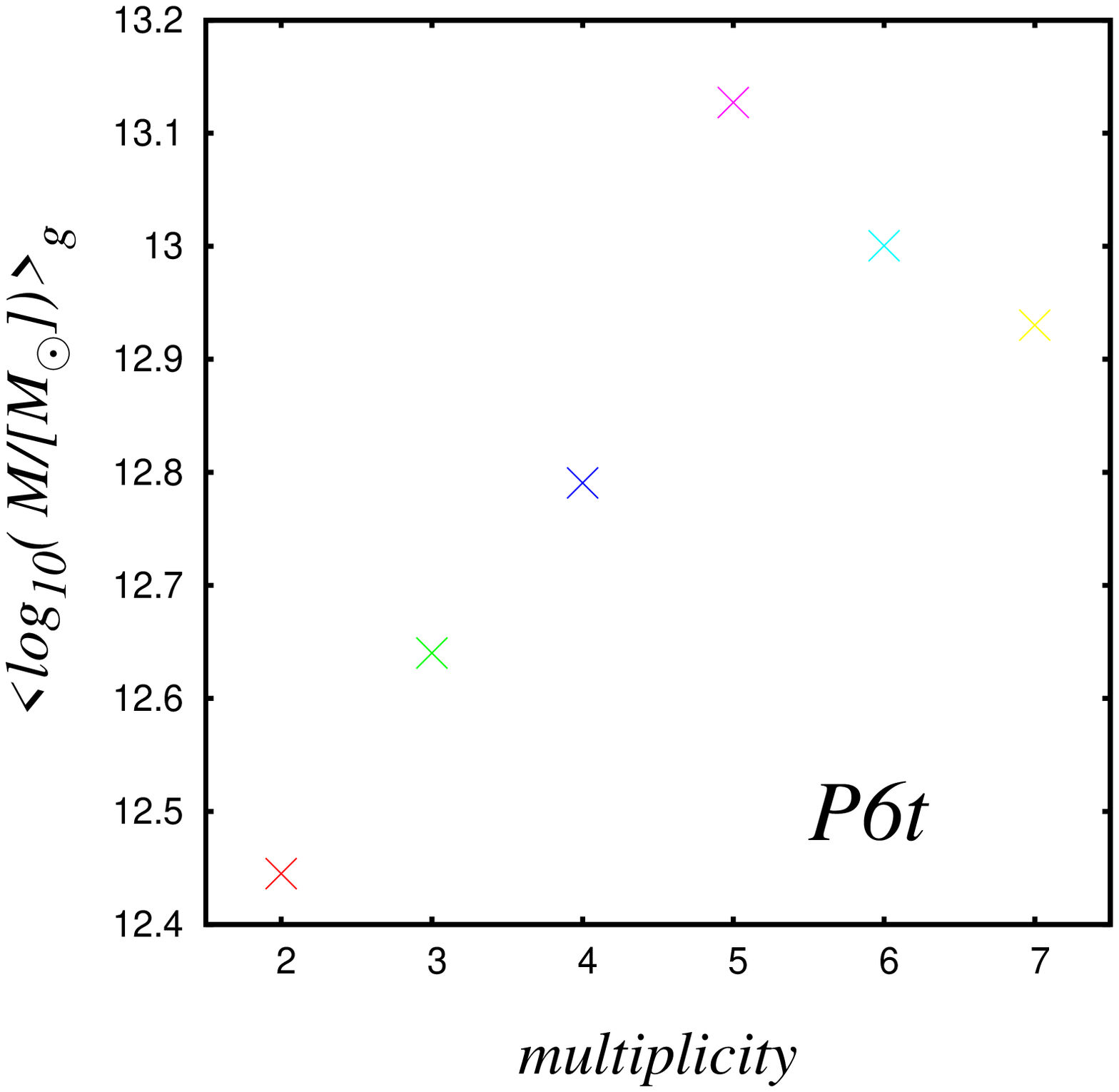} & \hspace{-1 cm} \includegraphics[width=3 in]{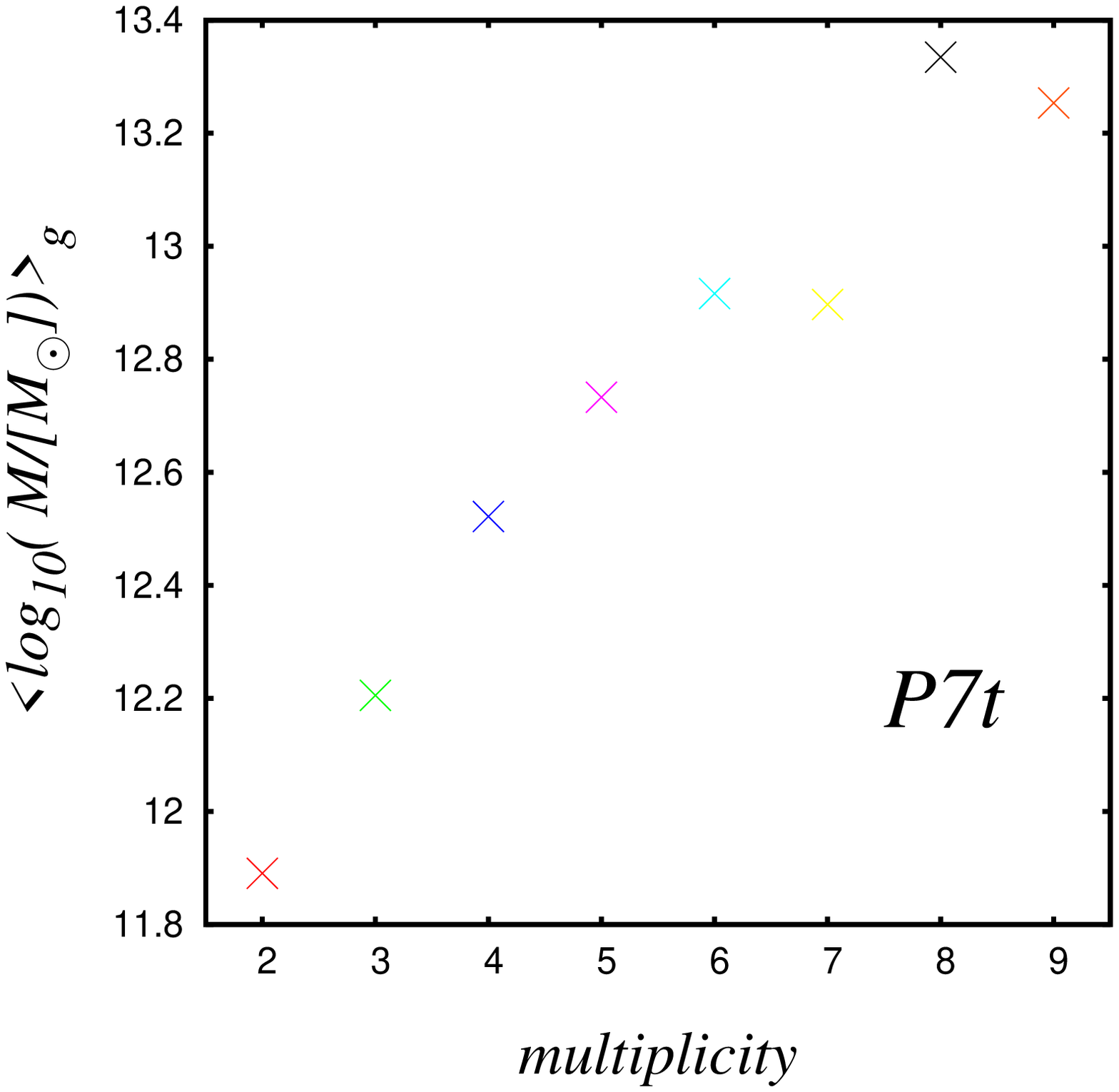}
\end{tabular}
\caption{\label{Fig:PropGru_Masa} The average mass of the groups detected in terms of the multiplicity shown in the x-axis. A 
line is included in the top-right panel to fit the data with the following 
formula: $lm_f(x) = lm_{a_1}*(x/m_s)+lm_{a_0}$. The values of the fitting parameters 
are $lm_{a_1}=0.29483$, $lm_{a_0}=12.1551$ and $m_s=1.78779$.}
\end{center}
\end{figure}

By only taking into account the groups of gas clumps detected after these conditions, we now 
calculate their average physical properties in terms of both their multiplicity and the 
partition in which they were found. The average mass resulting of the groups  
is shown in Fig.\ref{Fig:PropGru_Masa}. According to Section 
\ref{sec:sim}, the gas particles of the simulation have all the same mass, as given by 
$m_p \approx 6.4 \times 10^{8} M_{\odot}$, so that the plots of Fig.\ref{Fig:PropGru_Masa}  
also contain information about the average number of gas 
particles per group of a given multiplicity. 

%%%%%%%%%%%%%%%%%%%%
%%%%%%%%%%%%%%%%%%%%%

The panels of Fig.\ref{Fig:PropGru_Masa} seem to indicate 
that the mass of a middle multiplicity group is always larger than the mass of the highest 
multiplicity group detected.  

\begin{figure}
\begin{center}
\begin{tabular}{cc}
\includegraphics[width=3 in]{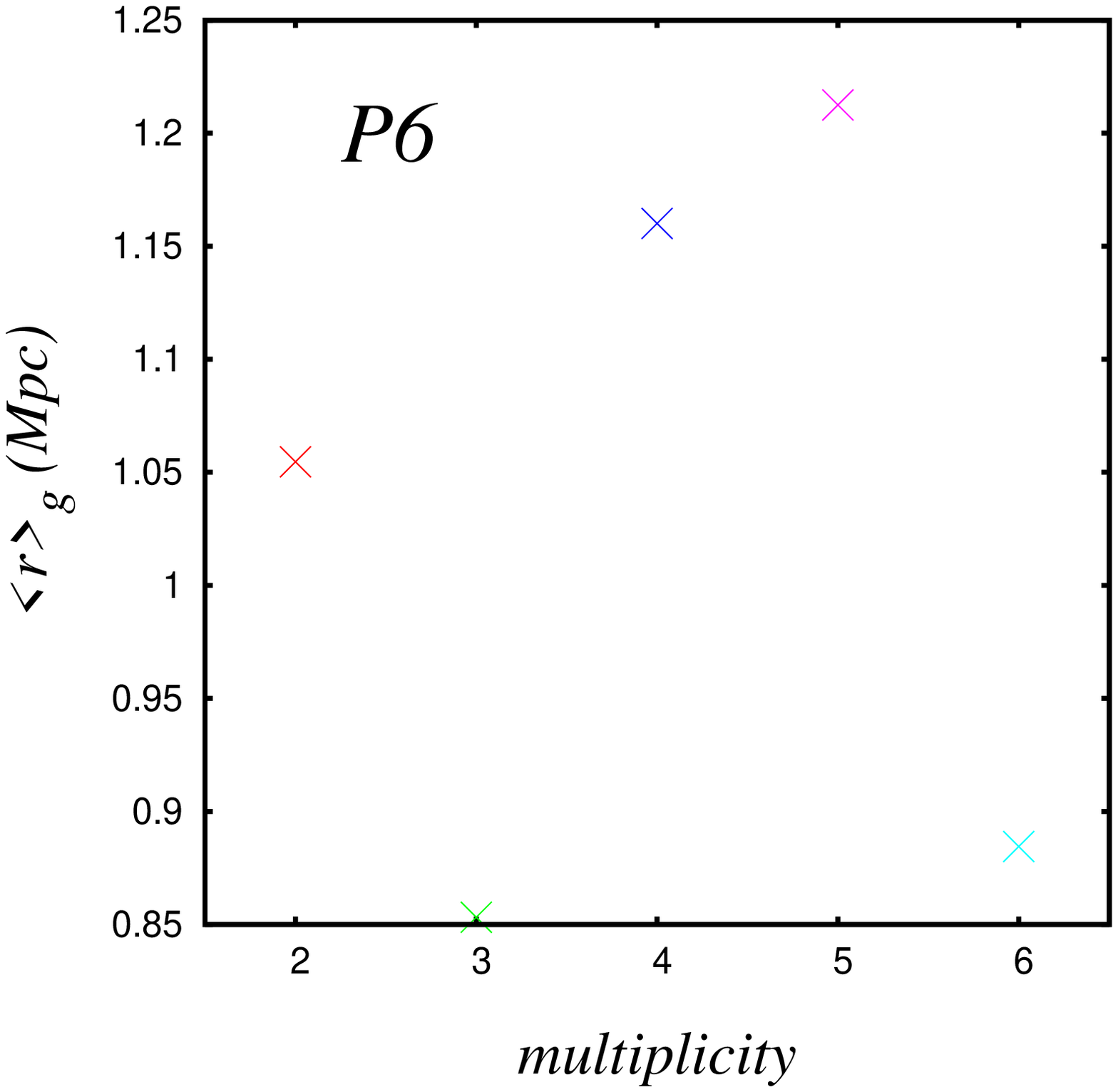} & \hspace{-1 cm} \includegraphics[width=3 in]{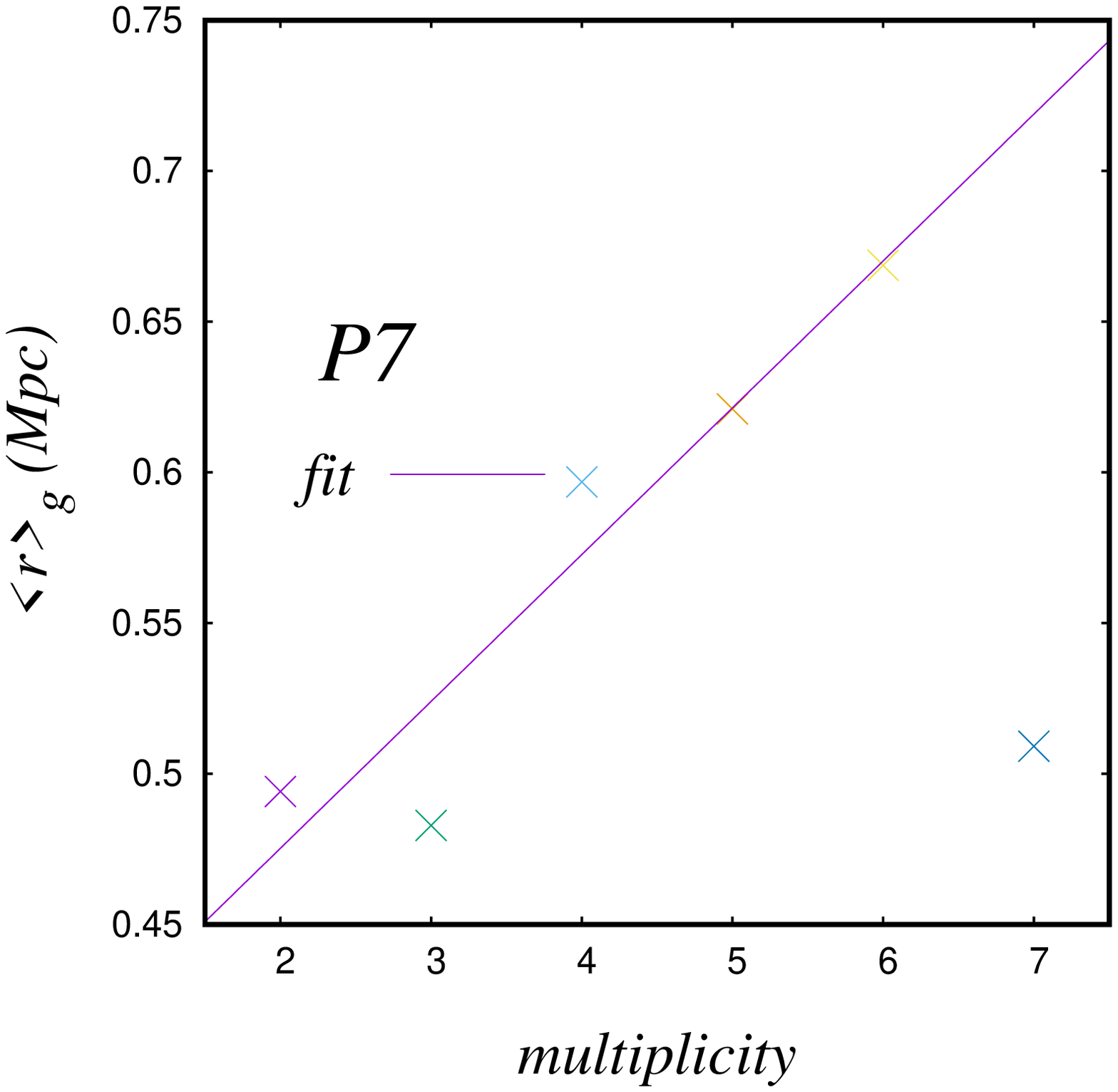}\\
\includegraphics[width=3 in]{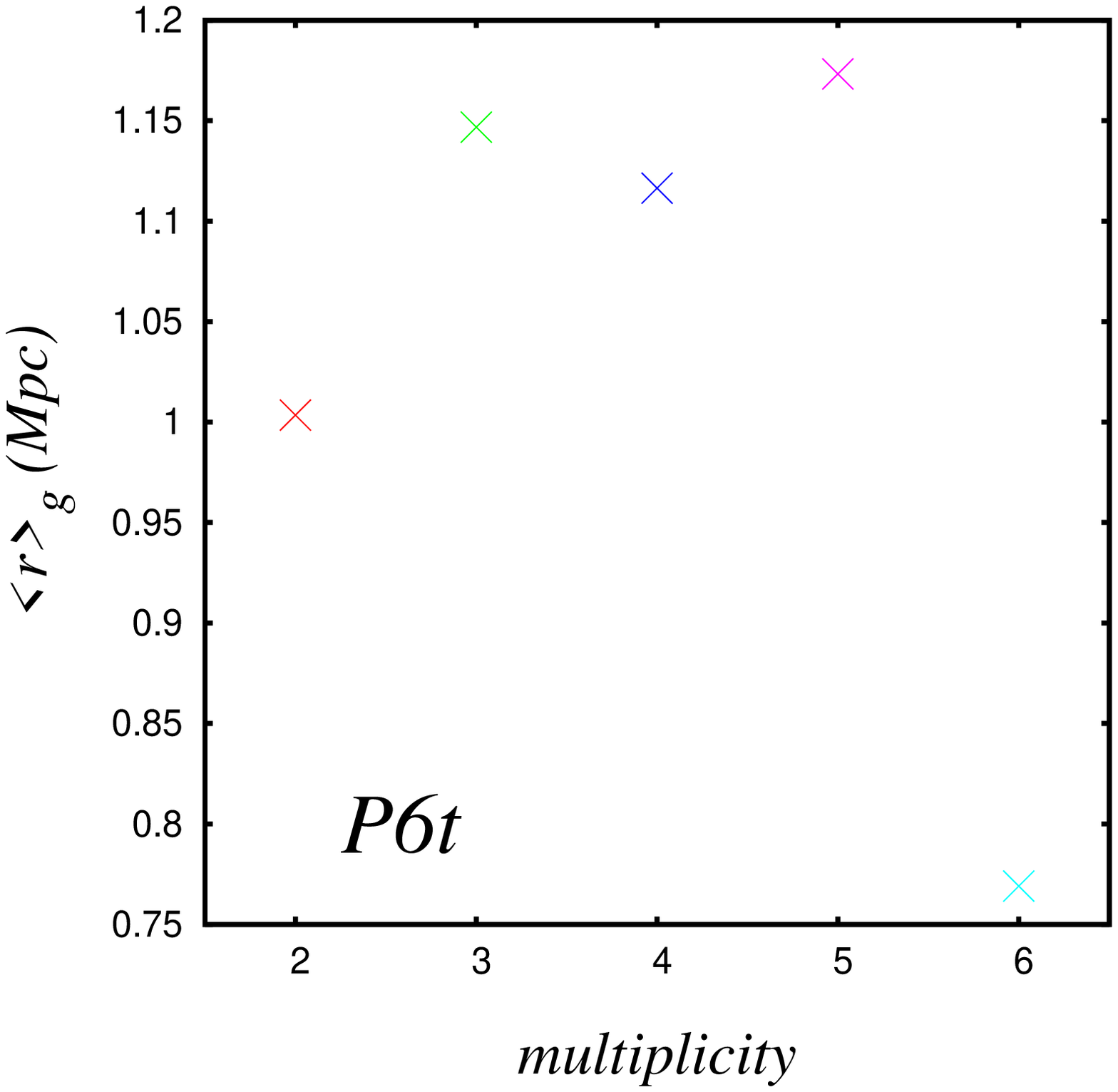} & \hspace{-1 cm} \includegraphics[width=3 in]{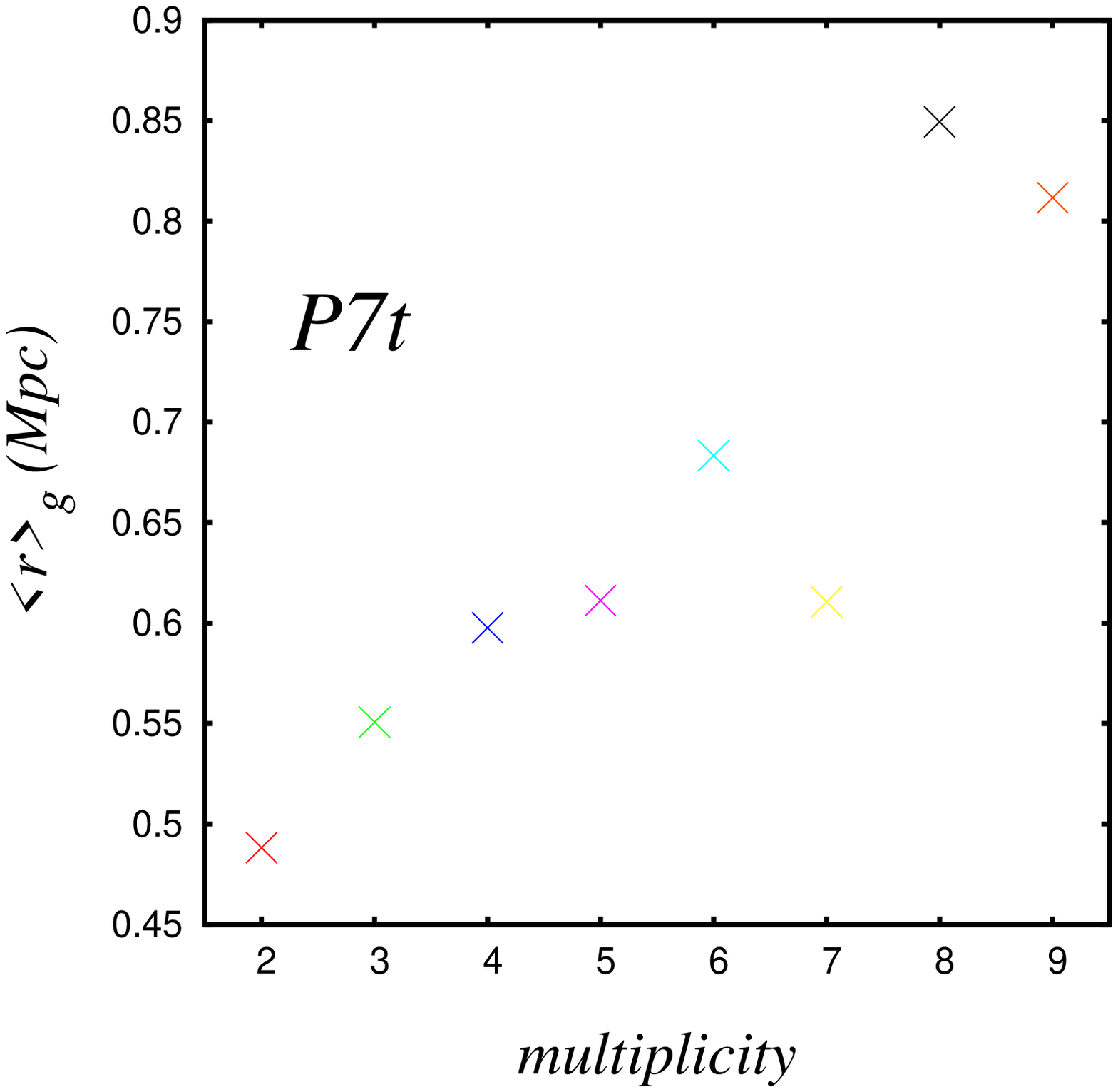}
\end{tabular}
\caption{\label{Fig:PropGru_Rad} The average radius of the groups detected with the multiplicity shown in the x-axis. A 
line is included in the top-right panel to fit the data with the following 
formula: $rad_f(x) = rad_{1}*(x/m_s)+rada_{0}$. The values of the fitting parameters 
are $rad_1=0.0795668$, $rad_0=0.37783$ and $m_s=1.63$.}
\end{center}
\end{figure}

The average radius per group is shown in Fig.\ref{Fig:PropGru_Rad}. It should be emphasized that 
this radius makes sense only in geometrical terms. To obtain this radius, we first calculate the 
center of mass of the set of gas particles per each group. Then, we determine the gas particle 
furthest away from this point and take half of this distance as the radius per each 
group. Next, we calculate the average of all these radii and we show the result in terms of 
the group multiplicity. The panels of Fig.\ref{Fig:PropGru_Rad} indicate 
that the radius of a lower multiplicity group is larger than the radius 
of the highest multiplicity group detected.

%%%%%%%%%%%%%%%%
%%%%%%%%%%%%%%%%   

\begin{figure}
\begin{center}
\begin{tabular}{cc}
\includegraphics[width=3 in]{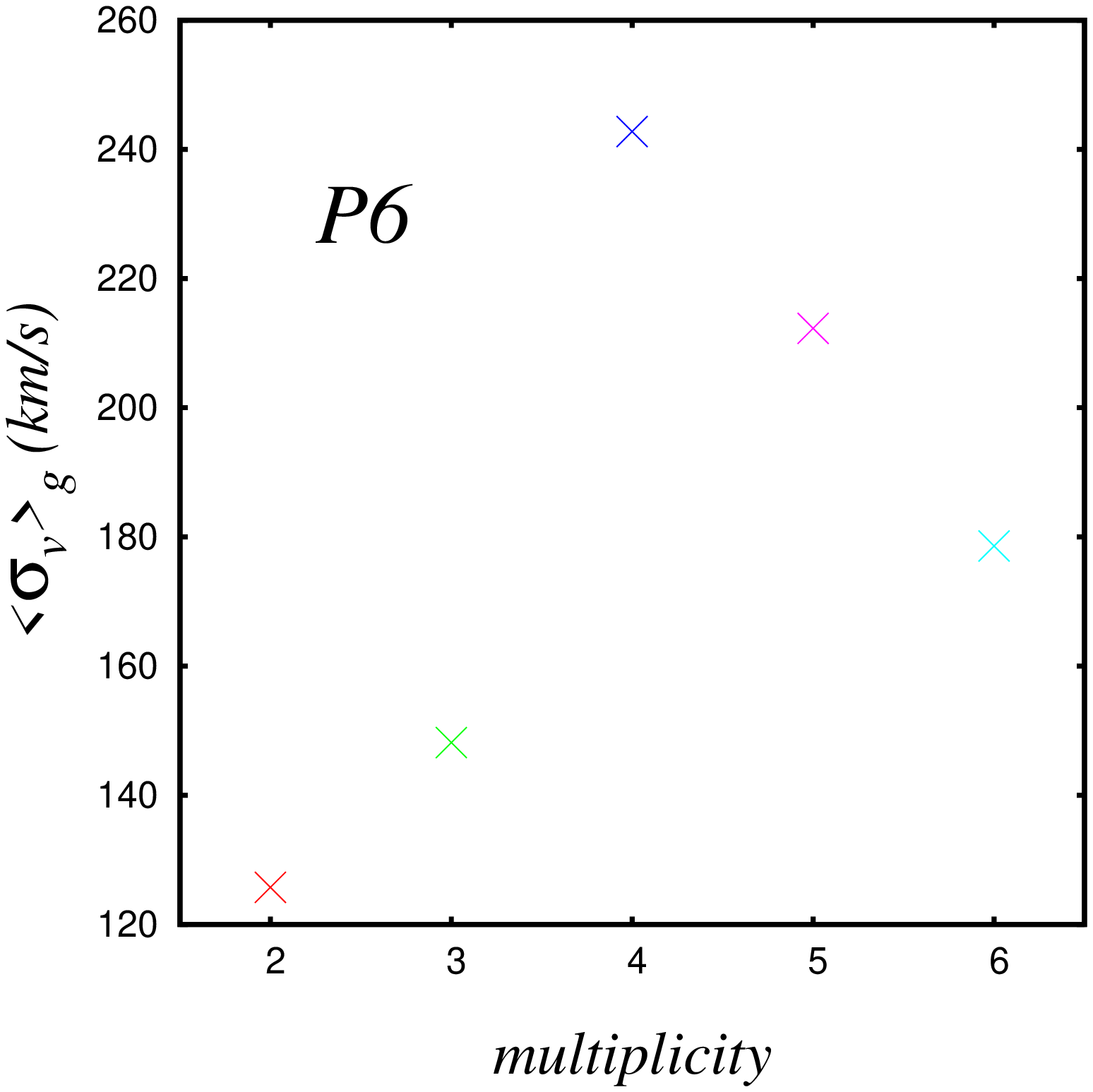} & \hspace{-1 cm} \includegraphics[width=3 in]{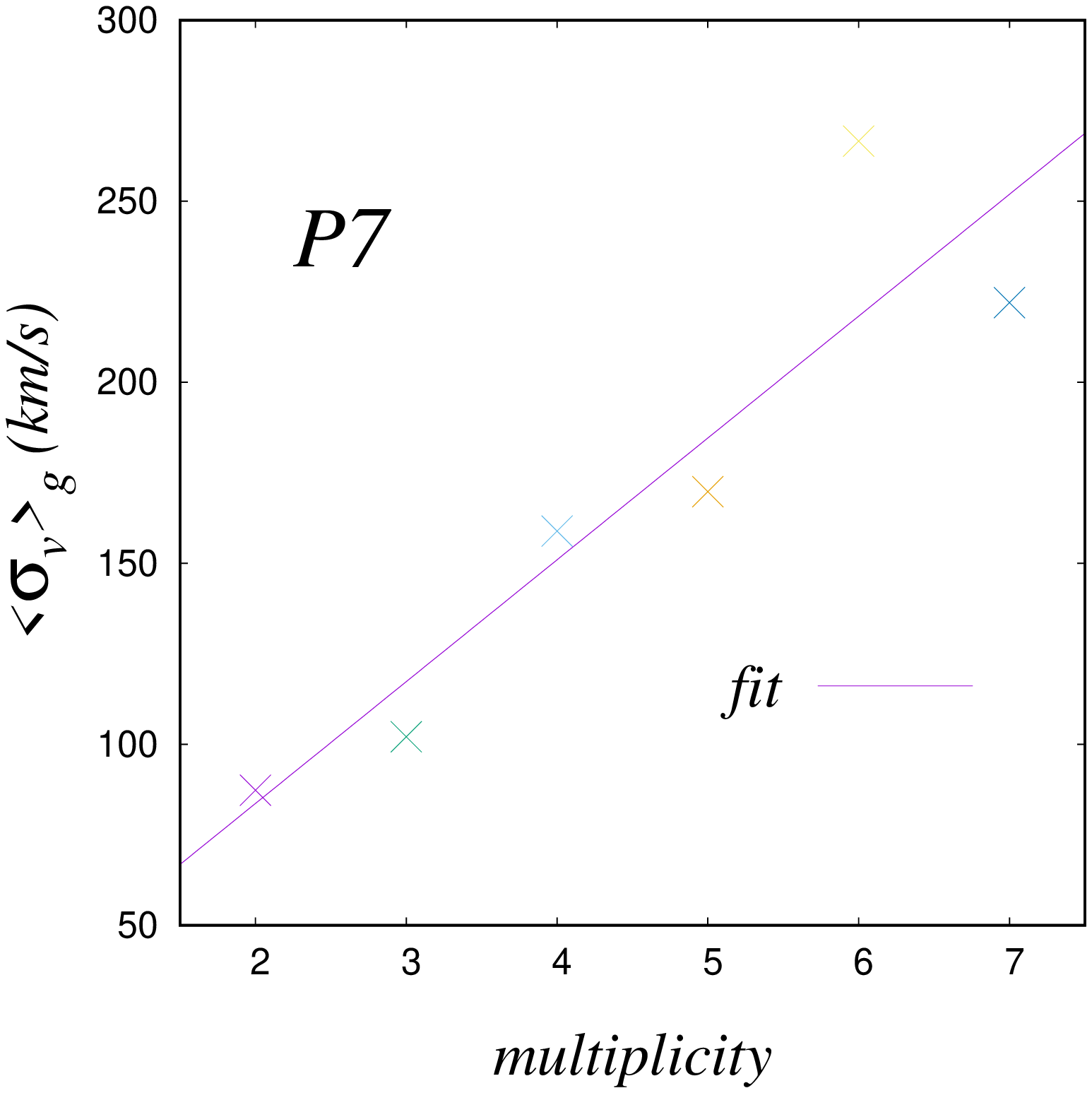}\\
\includegraphics[width=3 in]{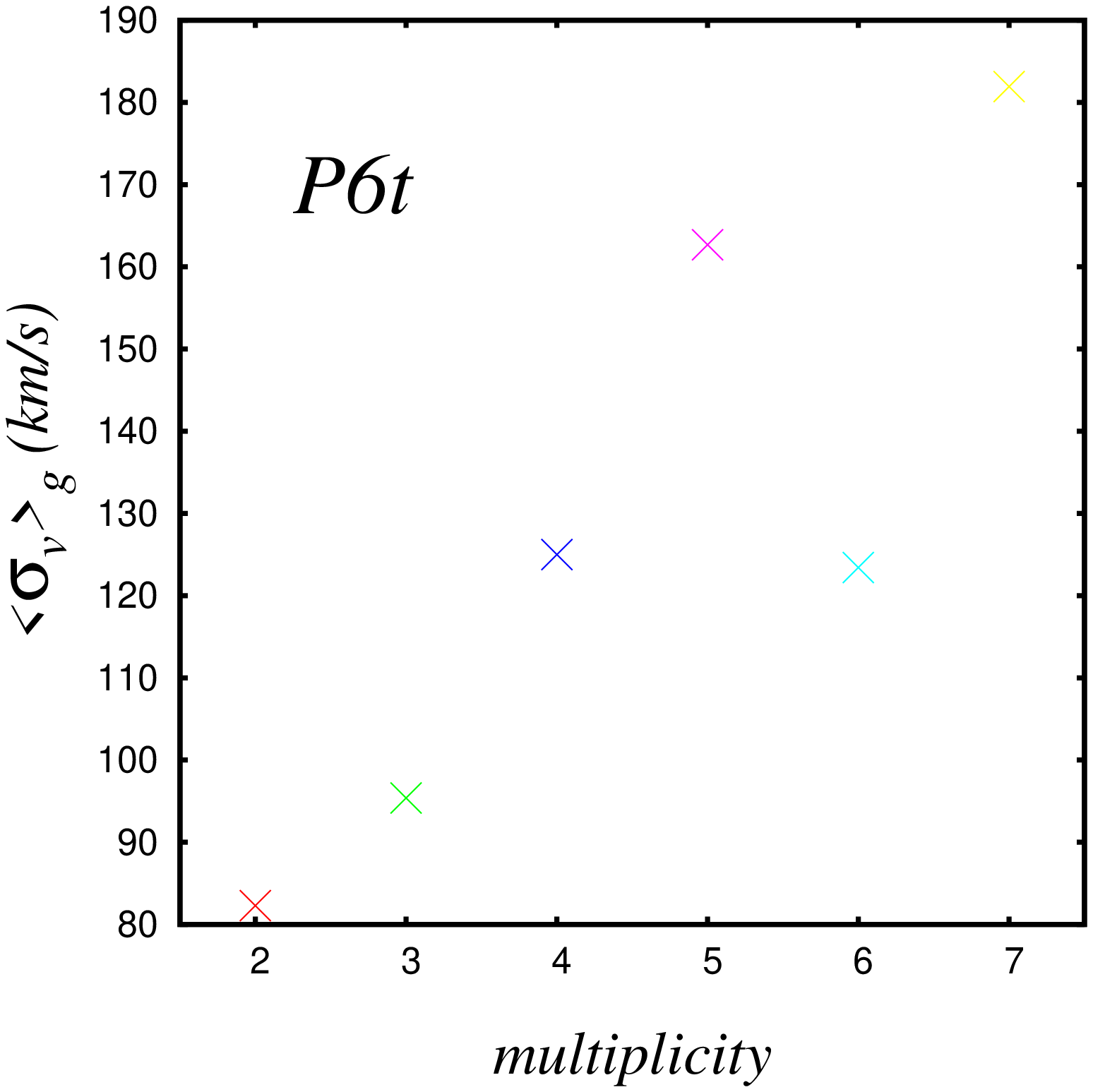} & \hspace{-1 cm} \includegraphics[width=3 in]{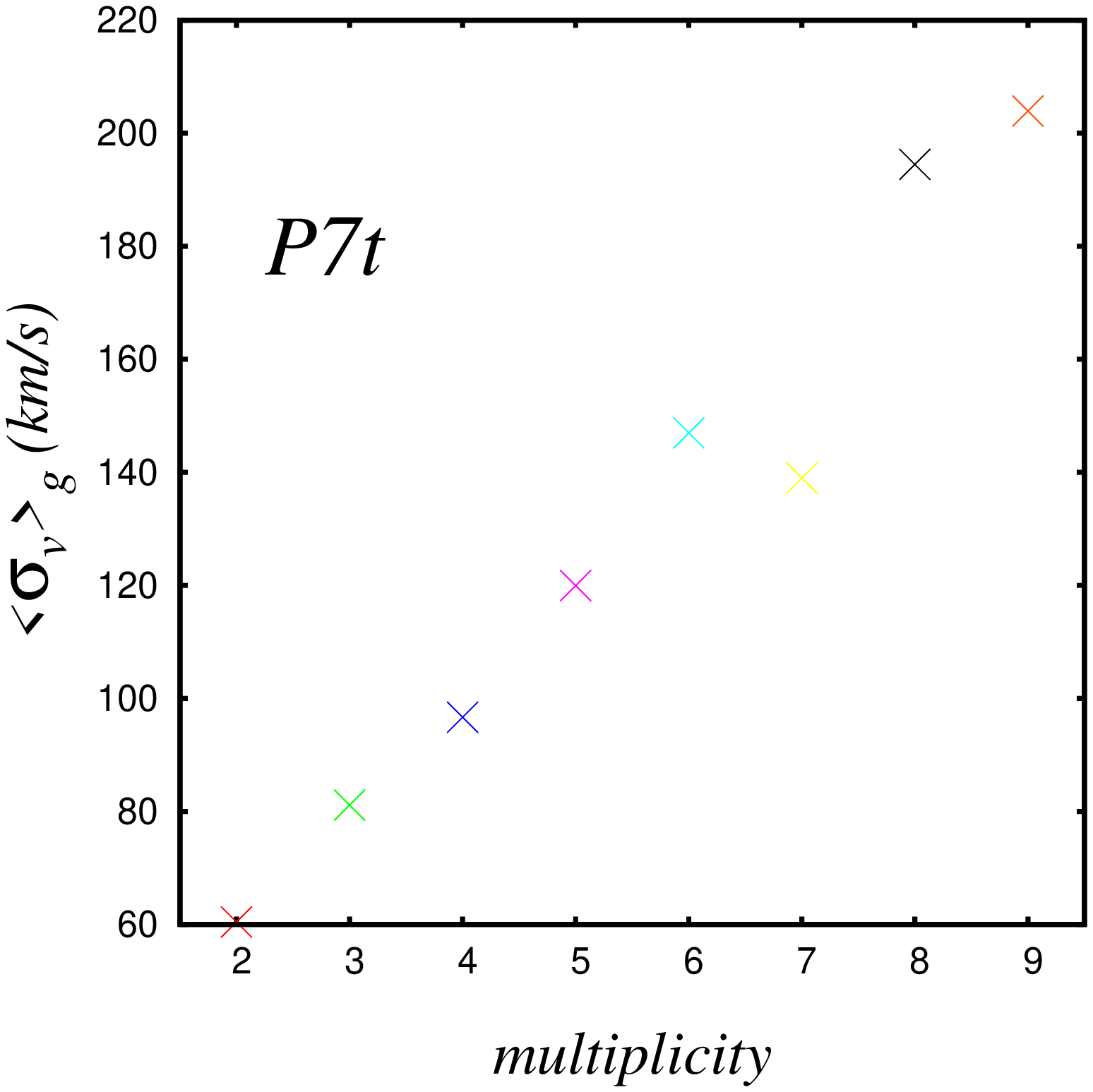}
\end{tabular}
\caption{\label{Fig:PropGru_SigmaVel} The average velocity dispersion of the groups detected in terms of the multiplicity shown in the x-axis. A 
line is included in the top-right panel to fit the data with the following 
formula: $\sigma_f(x) = \sigma_{1}*(x/m_s)+\sigma_{0}$. The values of the fitting parameters 
are $\sigma_1=1.47406$, $\sigma_0=16.326$ and $m_s=0.0437996$.}
\end{center}
\end{figure}

Finally, the average velocity dispersion of the groups detected is shown in Fig.\ref{Fig:PropGru_SigmaVel}. We 
emphasize that this velocity dispersion is calculated in the usual statistical sense: we calculate the magnitude of 
the velocity vector for all the gas particles of a given group, we then obtain the average 
velocity of all the groups and the dispersion is determined as the standard deviation, so that it is the 
square root of the variance; see \cite{numrecip}. The results are shown in terms of the multiplicity number. 

%%%%%%%%%%%%%%%
%%%%%%%%%%%%%%

\begin{figure}
\begin{center}
\begin{tabular}{cc}
\includegraphics[width=3 in]{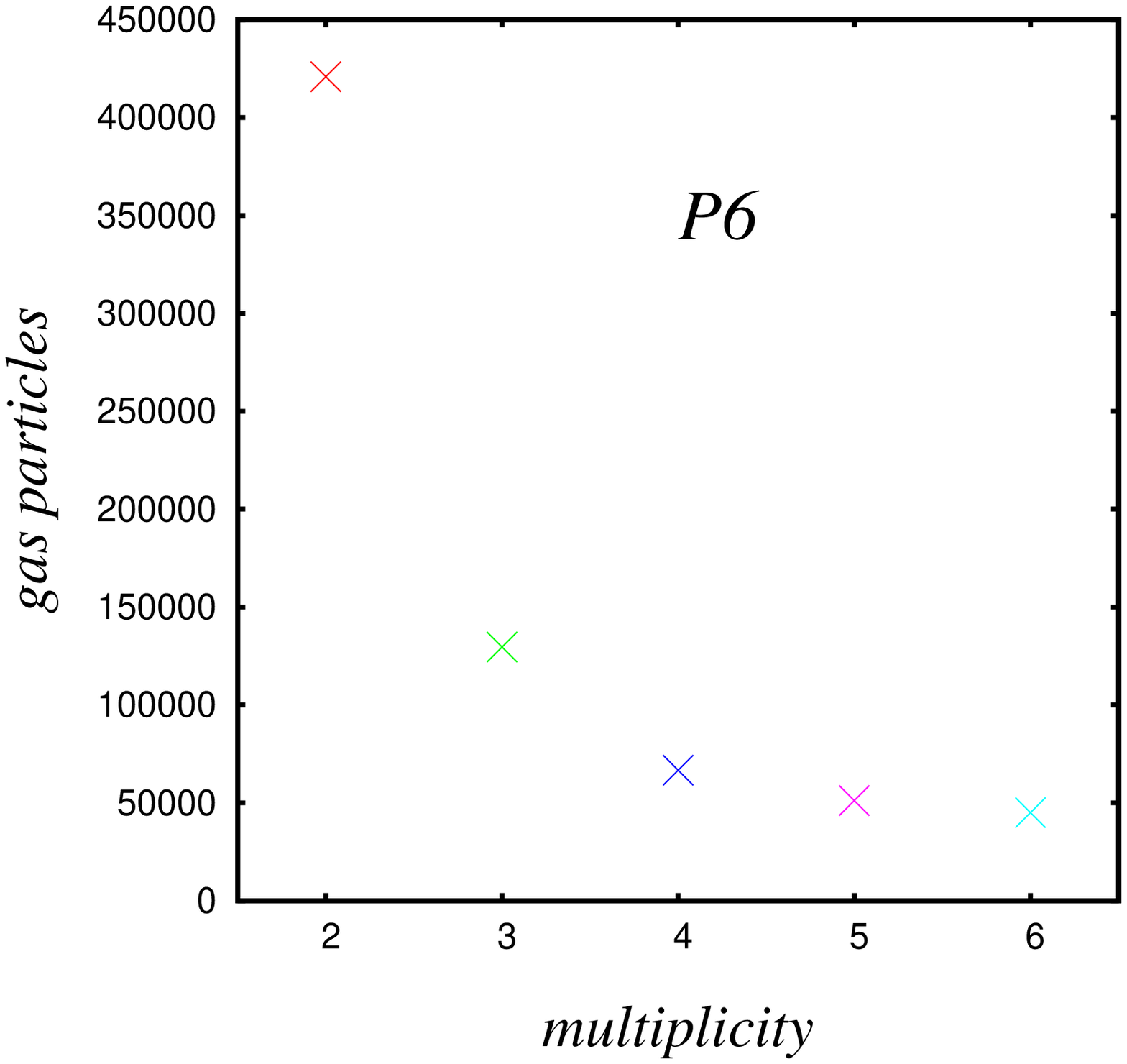} & \hspace{-1 cm} \includegraphics[width=3 in]{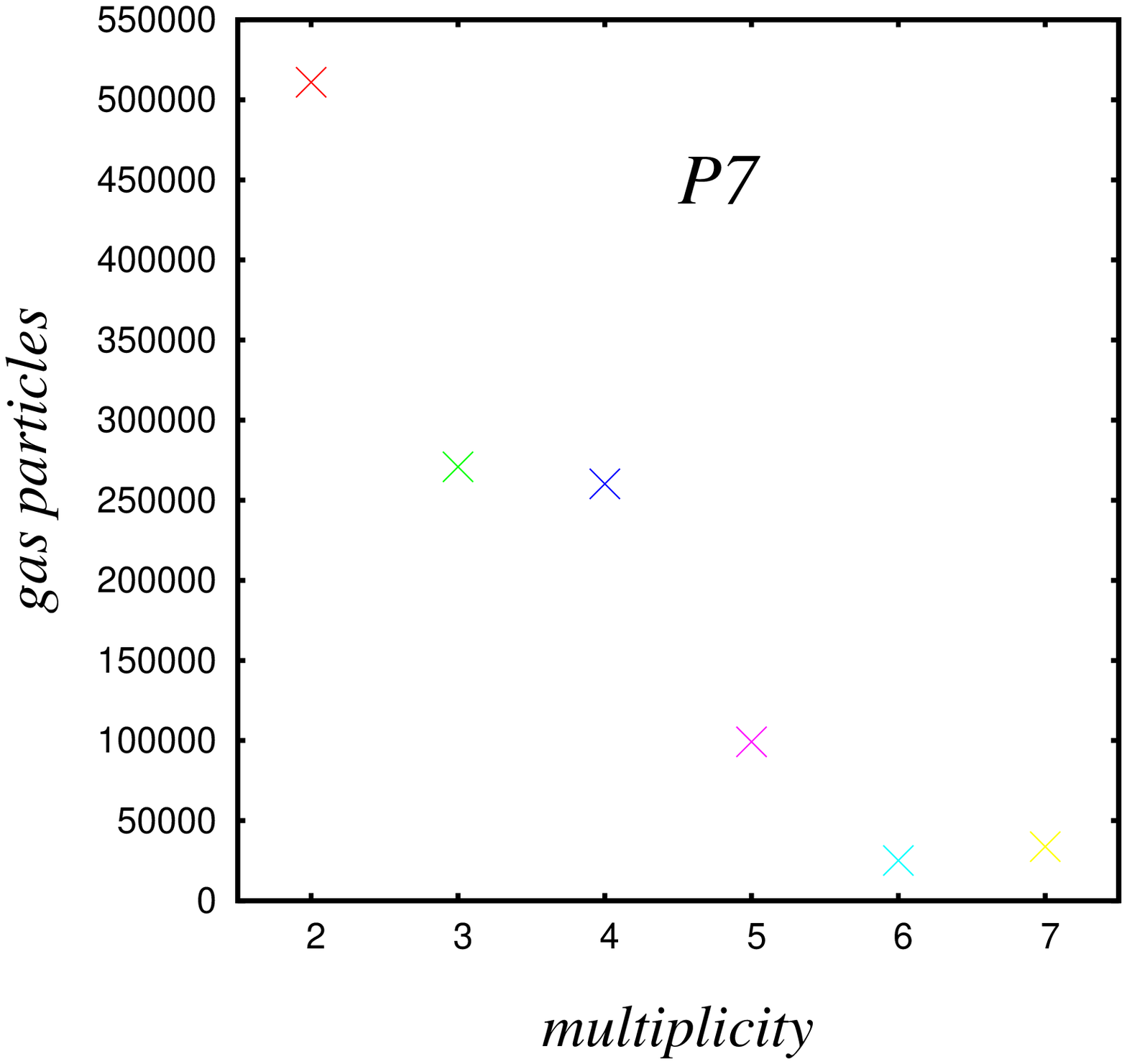} \\
\includegraphics[width=3 in]{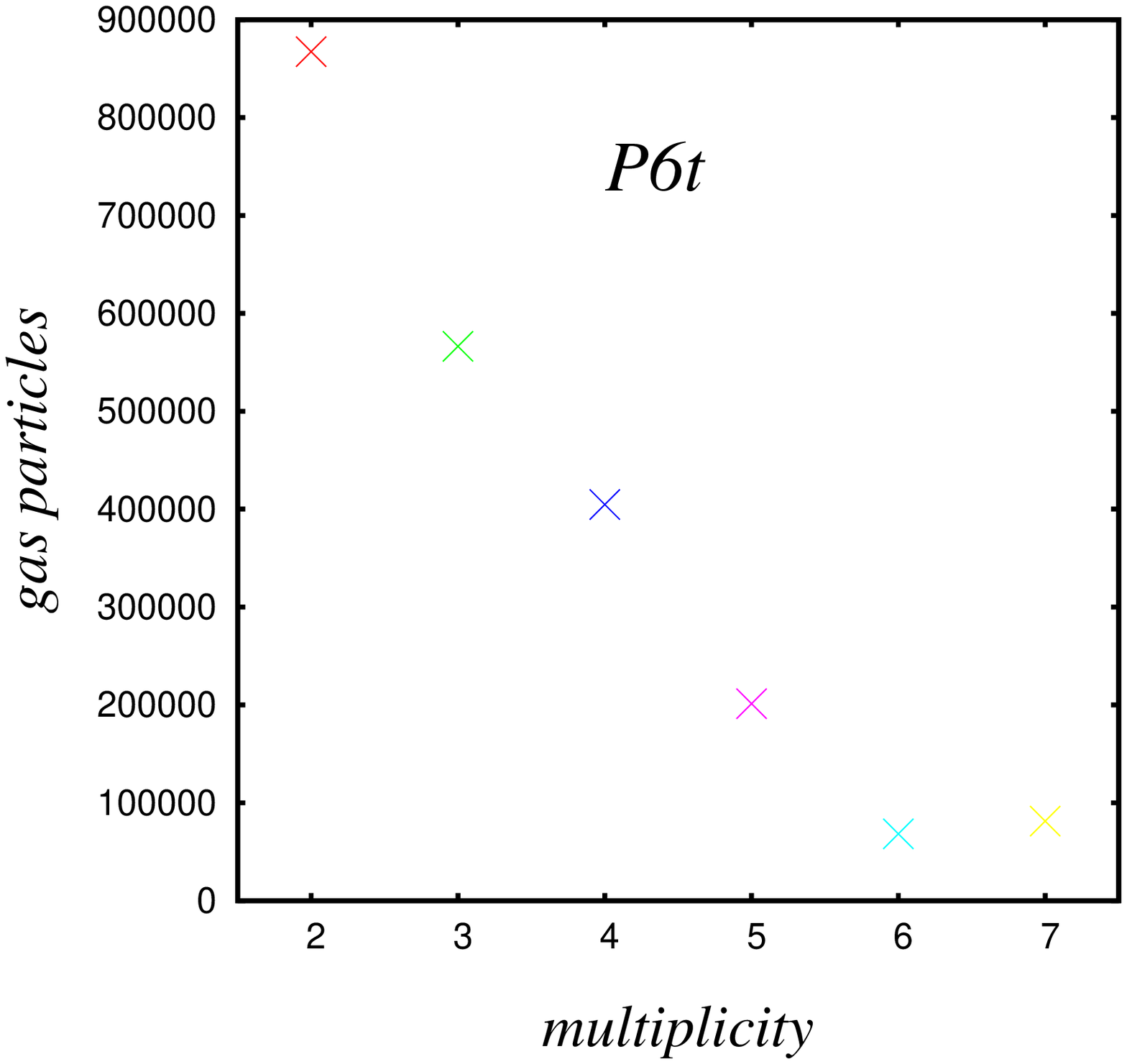} & \hspace{-1 cm} \includegraphics[width=3 in]{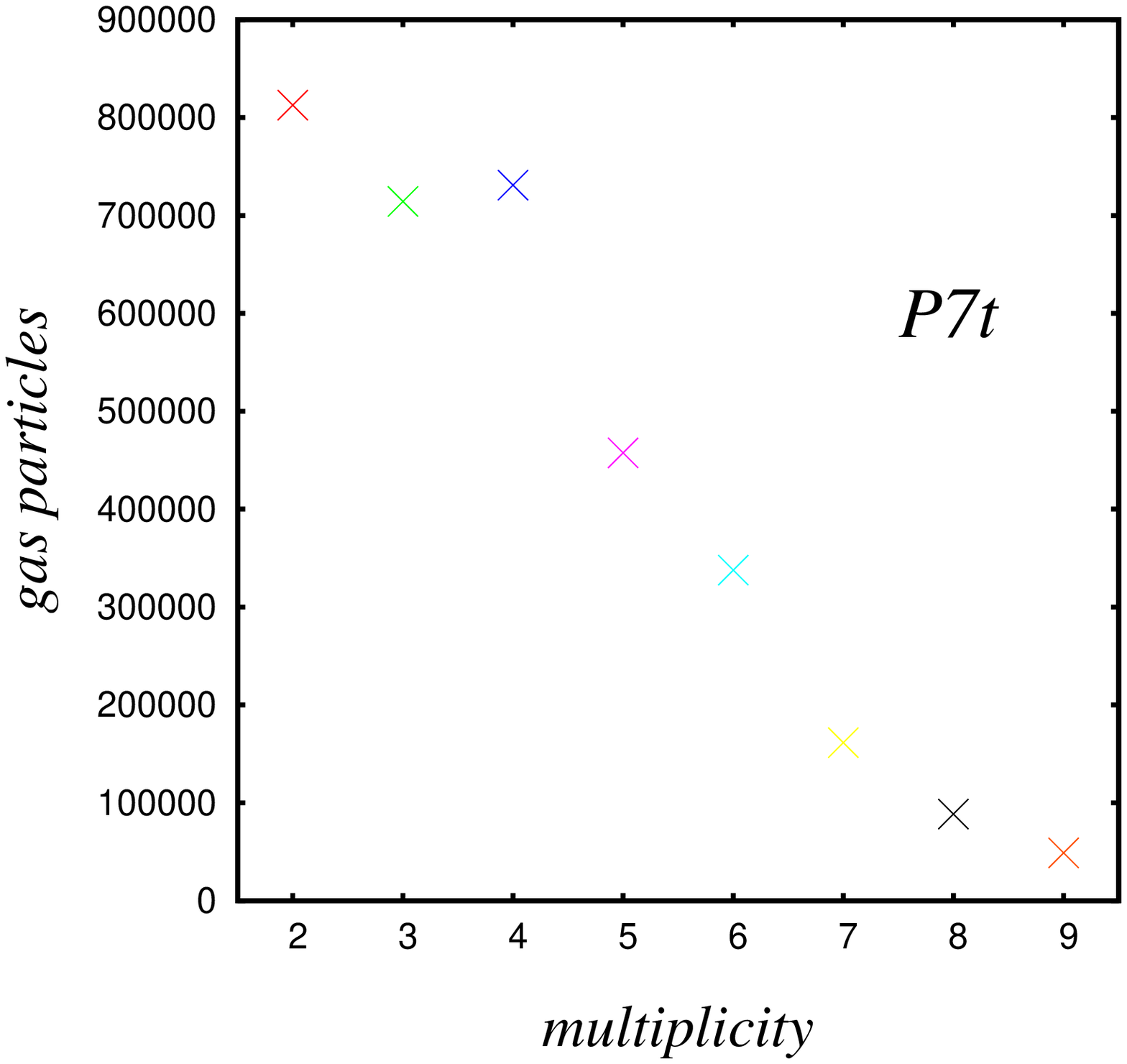}
\end{tabular}
\caption{\label{Fig:PropPart_Cont} The total number of gas particles found in any group with the multiplicity shown in the x-axis.}
\end{center}
\end{figure}

It should be emphasized the use of the symbol $< >_g$ (in the vertical axis of all of the figures described 
up to this point) indicates that the physical properties are calculated using all of the gas particles associated to 
all of the detected groups. However, to complement Table \ref{tab:stat}, we 
now consider all the gas particles of a given multiplicity but this time irrespective of the particular group 
to which they belong. The total number of gas particles detected in terms of their multiplicity is shown 
in Fig.\ref{Fig:PropPart_Cont}.

%%%%%%%%%%%%%%%%%
%%%%%%%%%%%%%%%%%%%%%%%

To take advantage of these results, we now summarize in Table \ref{tab:fitingf}, the fitting curves for all of 
the physical properties of the galaxy groups, that is, those indicated previously by the symbol $< >_g$.

A measure of the variation of the fitted data can be obtained by calculating the sum of squares of residuals, denoted here 
by $SS$, so that a standard error can be defined by $SE=\sqrt{SS/dF}$, where $dF$ is the number of degrees of freedom, which 
is defined as the number of fitted points minus the number of fitting parameters. The goodness of a fit can be
established by introducing the variance of residuals (or reduced $\chi_{\nu}^{2}$), which is given 
by $SE^2$. These values are reported in the last column of Table \ref{tab:fitingf} for each fitting formula. In principle,
a value of $\chi_{\nu}^{2}$ around 1 indicates a good fit; a value of $\chi_{\nu}^{2} \gg 1$ indicates a poor model 
fit; finally, a value of $\chi_{\nu}^{2} < 1$ indicates that there is noise in the model fit, which is usually named 
as "over-fitting" the data.

The goodness of the fitting parameters can be established by a confidence interval, which is determined by the best value 
of a fitting parameter $\pm \; tS*SE$, where $tS$ is the value from the $t$-distribution for the specified number of degrees 
of freedom and with 95 percent of confidence. In Table \ref{tab:confidence} we show these confidence intervals, so that the order 
of the fitting formulae shown in Table \ref{tab:fitingf} is still followed in Table \ref{tab:confidence}.

Finally, by calculating the mean $x_{\rm ave}$ and standard deviation $SD$ of the data array $x$, with $n_p=6$ points in this case, which are 
plotted in the vertical axis of each of the top right panels of Figures \ref{Fig:PropGru_Contf}-\ref{Fig:PropGru_SigmaVel}, a confidence 
interval of the physical properties can be determined by the equation $x_{\rm left}=x_{\rm ave}-tS*SD/\sqrt{n_p}$ for the left side 
and by the equation $x_{\rm right}=x_{\rm ave}+tS*SD/\sqrt{n_p}$ for the right side, so 
that these values are reported in columns 4 and 5 of Table \ref{tab:confidencephysicalP7}.

In order to quantify the influence that the use of the grid P6 can have on the results, in Table \ref{tab:confidencephysicalP6} we repeat 
only the calculation described in the paragraph above, by using the physical properties plotted in the vertical axis of each of the top 
left panels of Figures \ref{Fig:PropGru_Contf}-\ref{Fig:PropGru_SigmaVel}.

%%%%%%%%%%%%%%%%%%%%%%%%%%%%%%%%%%%%%%%%%%%%%%%%%%%%%%%%%%%%%%%%%%%%%%%%%%%%%%%%%%%%
\begin{table}[ph]
\caption{Fitting formulae of the mean physical properties of groups in terms of their multiplicity (m)}
{\begin{tabular}{|c|c|c|c|} 
\hline
\hline
symbol                                 & fitting formula                                      & Figure                         &  variance of residuals $\chi_{\nu}^{2}$  \\
\hline
\hline
$N_g(m)$                               & $1802.23 \times \exp \left(-m/0.841847 \right)$      & Fig.\ref{Fig:PropGru_Contf}    &        8.37           \\
\hline
$N_g(m)$                               & $17.1243 \times \left(m/4.09685 \right)^{-3.18728}$  & Fig.\ref{Fig:PropGru_Contf}    &       10.94            \\
\hline
$<\log \left( M/M_{\odot} \right)>(m)$ & $0.29483 \times (m/1.78779)+ 12.1551$                & Fig.\ref{Fig:PropGru_Masa}     &        0.04              \\
\hline
$<r>(m)$                               & $0.0795668 \times (m/1.63) + 0.37783$                & Fig.\ref{Fig:PropGru_Rad}      &        0.001             \\
\hline
$<\sigma_v>(m)$                        & $1.47406 \times (m/0.0437996) + 16.326$              & Fig.\ref{Fig:PropGru_SigmaVel} &     1253.93                \\
\hline
\hline
\end{tabular} }
\label{tab:fitingf}
\end{table}
%%%%%%%%%%%%%%%%%%%%%%%%%%%%%%%%%%%%%%%%%%%%%%%%%%%%%%%%%%%%%%%%%%%%%%%%%%%%%%%%%%%%%%%%%
\begin{table}[ph]
\caption{Confidence interval of the best value of the fitting parameter. The order 
of the fitting formulae shown in Table \ref{tab:fitingf} is still followed here}
{\begin{tabular}{|c|c|c|c|}
\hline
symbol & best value  &  left side    & right side  \\
\hline
\hline 
$I_1$  & 1802.23     &  1796.1  & 1808.4  \\
$m_s$  & 0.841847    &  -5.3    & 7.0     \\
\hline
$I_2$  & 17.12       &  9.3     & 24.9    \\     
$n_2$  & 3.2         & -4.7     & 11.0    \\
$m_{s2}$ & 4.09        & -3.8     & 11.9    \\  
\hline 
$lm_{a_1}$ & 0.29        & -0.17    & 0.76    \\ 
$lm_{a_0}$ & 12.16       & 11.67    & 12.62   \\          
$m_s$      & 1.79        & 1.32     & 2.26    \\
\hline
$rad_1$ & 0.079       & -0.026   & 0.18    \\
$rad_0$ & 0.378       &  0.27    & 0.49    \\
$m_s$   & 1.63        &  1.53    & 1.74    \\ 
\hline
$\sigma_1$ & 1.47        & -81.84   & 84.79   \\
$\sigma_0$ & 16.326      & -67.0    & 99.6    \\  
$m_s$      & 0.0437996   & -83.3    & 83.3    \\
\hline
\hline
\end{tabular} }
\label{tab:confidence}
\end{table}
%%%%%%%%%%%%%%%%%%%%%%%%%%%%%%%%%%%%%%%%%%%%%%%%%%%%%%%%%%%%%%%%%%%%%%%%%%%%%%%%%%%%%%%%%%%%%%%%%%%%
\begin{table}[ph]
\caption{Confidence interval of the physical properties by using the grid partition P7}
{\begin{tabular}{|c|c|c|c|c|}
\hline
\hline
Label                             &  mean      & standard deviation   &    left side  &  right side \\
\hline
$N_g$                             &  40.8 &    59.11         &    -6.47      &  88.13   \\
\hline
$<\log \left( M/M_{\odot} \right)>$ &  12.9 &     0.31         &    12.65      &  13.15  \\
\hline
$<r> (Mpc) $                      &   0.56 &     0.07         &     0.50      &   0.62   \\
\hline
$<\sigma_v> (Km/s)$               & 167.78 &    62.69         &     117.60    &  217.94  \\
\hline
\hline
\end{tabular} }
\label{tab:confidencephysicalP7}
\end{table}
%%%%%%%%%%%%%%%%%%%%%%%%%%%%%%%%%%%%%%%%%%%%%%%%%%%%%%%%%%%%%%%%%%%%%%%%%%%%%%%%%%%%%%%%%%%%%%%%%%%%
%%%%%%%%%%%%%%%%%%%%%%%%%%%%%%%%%%%%%%%%%%%%%%%%%%%%%%%%%%%%%%%%%%%%%%%%%%%%%%%%%%%%%%%%%%%%%%%%%%%%
\begin{table}[ph]
\caption{Confidence interval of the physical properties by using the grid partition P6}
{\begin{tabular}{|c|c|c|c|c|}
\hline
\hline
Label                             &  mean      & standard deviation   &    left side  &  right side \\
\hline
$N_g$                             &  11.4      &    14.61             &    -1.77      &    24.57  \\
\hline
$<\log \left( M/M_{\odot} \right)>$ &  13.11     &     0.19             &     12.9      &    13.23 \\
\hline
$<r> (Mpc) $                      &  1.03      &     0.14             &     0.90      &     1.16  \\
\hline
$<\sigma_v> (Km/s)$               &  181.5     &    42.26             &     143.42    &    219.58 \\
\hline
\hline
\end{tabular} }
\label{tab:confidencephysicalP6}
\end{table}
%%%%%%%%%%%%%%%%%%%%%%%%%%%%%%%%%%%%%%%%%%%%%%%%%%%%%%%%%%%%%%%%%%%%%%%%%%%%%%%%%%%%%%%%%%%%%%%%%%%%
%%%%%%%%%%%%%%%%%%%%%%%%%%%%%%%%%%%%%%%%%%%%%%%%%%%%%%%%%%%%%%%%%%%%%%%%%%%%%%%%%%%%%%%%%%%%%%%%%%%%
\subsection{The distribution function of the ratio $\frac{E_{\rm kin} }{\left| E_{\rm grav}\right|}$ of the gas clumps}
\label{subs:betahalos}

We now continue the characterization of the dense gas clumps contained in the chosen cubic elements by calculating their ratio 
of the kinetic energy to the gravitational energy. The energies 
involved are:

\begin{equation}
\begin{array}{l}
E_{\rm ther}=\frac{3}{2}\sum_{i} \, m_{i}\frac{P_{i}}{\rho _{i}}\\
E_{\rm kin}=\frac{1}{2}\sum_{i} \, m_{i} v_i^{2},\\
E_{\rm grav}=\frac{1}{2}\sum_{i} \, m_{i}\Phi_{i}
\label{energiespart}
\end{array}
\end{equation}
\noindent where the summations
include all of the gas particles inside each chosen cubic element, so that $\Phi _{i}$ is the gravitational 
potential, $v_i$ is the velocity, $m_i$ is the mass and $P_i$ is the pressure.
 
In simulations of the gravitational collapse of clouds, the dynamic behavior is
mainly determined by the values of the ratio of the thermal energy
to the potential energy, $\frac{E_{\rm ther}}{\left|E_{\rm grav}\right|}$, and the ratio of the kinetic energy 
to the gravitational energy, $\frac{E_{\rm kin}}{\left|E_{\rm grav}\right|}$;
see \cite{miyama}. In addition, these ratios are very important to
determine the stability of a gas structure against
gravitational collapse, so that collapse and even fragmentation
criteria can be constructed in terms of these ratios; see
\cite{hachisu1}, \cite{hachisu2} and the references therein. These
ratios also characterize very well the gravitational collapse by
capturing the most representative events, including
fragmentation, which may leave an imprint on the value of these ratios; see
\cite{RMAA2012}.

It must be mentioned that most of the codes considered in \cite{halosmad}
apply a procedure to improve the list of particles that potentially belong to a halo.
We did not implement a complete procedure to remove unbound gas particles. However, only gas 
particles with $\Phi_{i}<0$ have been considered in the calculation of 
this Section~\ref{subs:betahalos}, because these
particles are gravitationally linked. Once these particles are selected
in each chosen cubic element, we then calculate the ratios defined in this section. It is important to emphasize 
that only those chosen cubic elements with a number of particles greater than 200 are considered in this calculation.  

\begin{figure}
\begin{center}
\includegraphics[width=4.0 in]{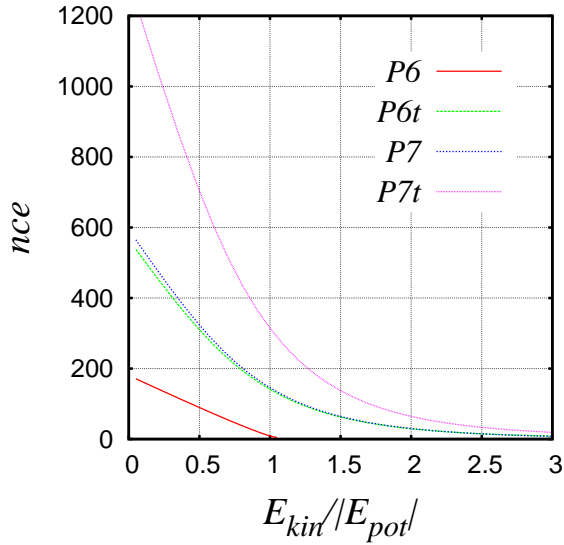}
\caption{\label{fig:FuncBeta} The distribution functions of the number of chosen cubic elements $nce$ (shown in the vertical axis)
against their ratio of kinetic energy to gravitational energy (shown in the horizontal axis).}
\end{center}
\end{figure}

We consider it to be more illustrative to present the results of this
calculation in terms of the number of cubic elements, $nce$ found within a
given interval of the ratio $\frac{E_{\rm kin}}{\left|E_{\rm grav}\right|}$, as has been done in 
previous sections the result is shown in Fig.~\ref{fig:FuncBeta}.

%%%%%%%%%%%%%%%%%%%%%%%
%%%%%%%%%%%%%%%%%%

The curves thus obtained indicate that most of the chosen cubic elements have 
low-values of the ratio $\frac{E_{\rm kin}}{\left|E_{\rm grav}\right|}$, so that most of the
gas inside each chosen cubic element is clustered such that its average value is 
$\frac{E_{\rm kin}}{\left|E_{\rm grav}\right|} < 1$.
%%%%%%%%%%%%%%%%%%%%%%%%%%%%%%%%%%%%%%%%%%%%%%%%%%%%%%%%%%%%%%%%%%%%%%%%%%%%%%%%%%%%%%%%%%%%%%%%%%%%%%%%%%%%%
%%%%%%%%%%%%%%%%%%%%%%%%%%%%%%%%%%%%%%%%%%%%%%%%%%%%%%%%%%%%%%%%%%%%%%%%%%%%%%%%%%%%%%%%%%%%%%%%%%%%%%%%%%%%%
\section{Discussion}
\label{sec:discu}

In this paper we applied a mesh-based code to generate a uniform cubic partition to characterize the 
clustering and grouping properties of gas clumps at galaxy cluster scales using a typical cosmological simulation. 

It must be first emphasized that the results reported in Sections~\ref{sec:results} 
depend strongly on the parameters of the partition, as expected. This is a common situation, even 
in highly-refined codes presented elsewhere, in which the main parameters of the code must be given before-hand. Here, we 
will consider other features to clarify the results obtained and their dependence on the partition used. The basic 
partition depends on two parameters, namely: the level of resolution and the density threshold. The 
low-resolution partitions were labeled as P6 and P6t, corresponding to a density 
threshold of 2 and 1.5, respectively; analogously, the high-resolution 
partitions were labeled as P7 and P7t, respectively.
 
A high fraction of the chosen cubic elements contained more than 500 particles, as can be seen 
in Fig.\ref{fig:FuncDistNpE}, so that the gas clumps are expected to be well represented in this 
low-resolution simulation. However, the edges of the chosen cubic elements 
of partition P6 and P6t were visible in Fig. \ref{EscoT_006}. This indicates that 
the size of their typical cell element was too big, and there can be more than one gas 
clump contained in each cell element. Fortunately, the edges were less visible in the 
partitions P7 and P7t, as can be seen in Fig. \ref{EscoT_007}. This indicates that only the core of 
the largest gas clumps were captured by each cell element. These features can be seen 
as shortcomings of this method. On the other hand, a positive feature of this method is 
that it can certainly be identified filamentary distributions of cubic elements. Nevertheless, by comparing 
the distributions of cubic elements seen in Figs. \ref{EscoT_006} and \ref{EscoT_007}, one then can see 
the critical role played by the normalized 
density threshold: the higher its value, the better its representation of the gas clumps by 
the set of chosen cubic elements.

A large number of chosen cubic elements were 
linked in lower-multiplicity groups. Specifically, in the case of multiplicity 2, the 
ratio of these numbers between the partitions 
P7 to P6 is around 8. This means that there are 8 times more binary systems of the chosen cubic 
elements in partitions P7 than those detected in partition P6. A similar ratio was observed in the case of 
partitions P7t to P6t; see Fig.\ref{fig:Mult}. This behavior indicates that a change in the resolution 
has a significant impact. Meanwhile, a small change in the density threshold does not affect too much 
the larger number of the chosen cubic elements detected in partitions P7 and P7t as compared with 
those obtained for partitions P6 and P6t. 

However, analogous ratios constructed with the number chosen cubic elements of multiplicity 2 between 
the partition P7t to P7 and P6t to P6 indicate that the change is smaller than that observed in the 
previous paragraph, so that it is now about 1.6. This means that the number of binary systems of chosen 
cubic elements does not duplicate when we change the density threshold from 2 to 1.5 in partitions with 
the same level of resolution. 

As noted in Section \ref{subs:grupos}, after imposing the two conditions on the gas particles contained in 
the chosen cubic elements, the number of groups detected in terms of their multiplicity decreased 
significantly. Let us consider again the case of 
multiplicity 2 that was discussed above, then the ratio between 
the numbers of groups detected in the partitions P7 to P6 and P7t to P6t are now of the order 
4. This means that now there are 4 times more binary systems of gas clumps in the partitions of level 7 than those 
detected in the partitions of level 6. Surprisingly, a comparison of the analogous ratios between the partitions 
P7t to P7 and P6t to P6 indicate that they are of the same order 4. 

The numbers of interest detected for partitions P6t and P7t almost doubled when compared to 
those detected for the original partitions P6 and P7 ( e.g. the number of chosen cubic elements, the 
number of groups, etc.) Meanwhile, the physical properties of the groups are comparable; for instance, the mass 
scale in terms of 
$\log_{10} \left( M/M_{\odot} \right)$ for partition P6 ranges within 12.9--13.4; while for partition P6t, it ranges 
within 12.4--13.1; for the partitions P7 and P7t we observed the normalized mass range within 12.3--13.3 
and 11.8--13.4, respectively.

Apparently, it will be easier to lose the gas clumps originally detected when 
the multiplicity of the associations of chosen cubic 
elements is higher. It is very likely that 
the gas clumps associated in higher-multiplicity groups are too small and do not meet the conditions 
imposed in Section \ref{subs:grupos}. It is also possible that 
the highest-multiplicity groups still detected in Section \ref{subs:grupos} are incomplete in their number 
of members in view of this behavior. 

While it is true that only gravitationally bounded gas particles were used to identify gas clumps and determine 
group properties in Section \ref{subs:grupos}, it must be noted that no test was made to check whether the gas clumps 
that were placed in groups are gravitationally bound to each other. In addition, the proximity parameter introduced 
in Section \ref{subs:funcionmult} was motivated by the size of the cubic element of the partitions, which 
means that it can be changed and the results are going to be changed accordingly. 

However, a proximity parameter can in principle be found, so that its cubic partition be the better choice for 
a given simulation. In fact, N-body simulations and the friends-of-friends algorithm 
have been used to calibrate these kind of codes to obtain values 
of linking length parameter, so that 
it makes the best identification of galaxy groups in catalogs; see for instance \cite{eke}, \cite{padilla} and \cite{nor}.  

In Section~\ref{subs:betahalos} we calculated
the ratio of kinetic energy to gravitational energy of all the gas 
clumps contained in the chosen cubic elements. These values can be useful as initial conditions for simulations of the formation 
of star clusters, like the ones simulated by \cite{klessen1} and \cite{klessen2}. However, the physical properties 
of the gas cloud progenitors that will produce the star clusters by gravitational collapse, are difficult 
to obtain, mostly because these cluster precursors are very difficult to be observed. Nevertheless, \cite{james} have 
recently observed "a high mass molecular cloud with unusually large linewidths", which indicate that its 
level of kinetic energy is so high, that this cloud was difficult to be considered as a gravitationally 
bound system. These authors claimed that if this system were successfully identified as a 
star cluster progenitor, then this cloud must be dominated by extreme turbulence. It has 
been shown elsewhere that this kind of highly turbulent cloud can still collapse and 
form protostars, see \cite{miass2018}.        
%%%%%%%%%%%%%%%%%%%%%%%%%%%%%%%%%%%%%%%%%%%%%%%%%%%%%%%%%%%%%%%%%%%%%%%%%%%%%%%%%%%%%%%%%%%%%%%%%%%%%%%%%%%%
\section{Comparison with other papers and observations}
\label{sec:comp}

A natural way to assess the results reported in this paper
is by means of a comparison with other simulations. In spite of the fact that our
gas structures are just a crude representation of the real distribution of galaxies, we will also try to make 
a comparison with observations. In order to keep these comparisons tractable, in this Section we will focus only 
on the results obtained by using the partition P7t, which has given us the best results. 

With regard to the numerical simulations side, there is a rich literature on 
multiplicity functions in cosmology that can be mentioned. For instance, using early dark-matter only 
simulatons, \cite{gott} described the clustering of galaxies in terms of a multiplicity
function, defined as the fraction of galaxies in single, double, triple systems, etc. \cite{bhavsar} determined 
the number of groups in terms of the number of galaxy members. 
\cite{efs} compared the multiplicity function obtained from N-body simulations with the 
Press and Schechter formalism, which presented an approximate theory for the form of the 
multiplicity function, see \cite{ps}. A mathematical definition of 
multiplicity is given by \cite{efs}, which is "the multiplicity of each particle 
is $m$ if it is part of a group with more than 2$^{m-1}$ but no more than 2$^m$ members". 

The ratio of the number of galaxies forming groups with 2 or 3 members 
between the number galaxies in groups of 4 or 5 members is 4 according to \cite{bhavsar} and 8 according to \cite{efs}.    
According with our plots shown in the right column of Fig.\ref{Fig:PropGru_Contf}, our ratio is around 8.  

\cite{thomas} presented simulations on the formation 
of a rich cluster of galaxies, in which many small galaxies are detected in the region around the central 
galaxy by using a minimum density cut of 180 times the mean density of the simulation, so that 12 gas clumps are located 
the outer region beyond 1 Mpc and 20 gas clumps within this radius. Figure 11 of \cite{thomas}, presented a projection 
to the X-Y plane of the particle distribution, it shows a galaxy distribution very similar to that shown here 
in Fig.\ref{fig:ElGrupo}. 

In addition, in the paper \cite{perez1}, the authors located the gravitational
bounded gas structures of a cosmological hydrodynamical simulation by using the friends-of-friends 
algorithm, see \cite{davis}. Then, they continued by looking for condensed gas substructures within 
regions of 0.5 Mpc, centered on each bounded system localized, so that they identified 
a galaxy-like object with those gas concentrations found as substructure. They found 364 systems in 
pairs. After applying a proximity criterion, they finally obtained 88 galaxies-like objects in pairs. From 
these sample they constructed a 3D catalog and made an statistical analisys. This method goes 
beyond the capabilities of the code presented in this paper, which does not allow to capture 
gas substructures, so that a comparison is dificult to make. However, their number of galaxies in pairs seems 
to be quite small as compared with the more than 700 binary systems we found by using the partition P7t.

\cite{salesa} found an average of about 10 satellite galaxies within the virial radius by using 
a friends-of-friends algorithm on a hydrodynamical simulation of a small region, which was 
taken from a cosmological simulation box and then re-simulated with the zoom-in technique, that is, at 
higher resolution but preserving the tidal fields from the whole box. It is interesting to mention that 
in this paper, groups of gas clumps up to 10 members were studied, because groups with a larger number of 
members were no found.

%Using the same cosmological simulation, \cite{salesb} considered several resimulations to follow 
%a single primary galaxy in detail and examine the orbits of its satellite galaxies, which appeared to be 
%smaller, gravitational bounded systems composed of stars, gas and dark matter. 

With regard to the observational side, it must be emphasized that catalogs of galaxies and 
cluster of galaxies have been generated from the data published by the 2dF Galaxy 
Redshif Survey, see for example \cite{tago2005} and \cite{tago2006}. The final release of the 2dFGRS contained 
245591 galaxies out of which all group of galaxies catalogues were constructed using selection criteria and 
a cluster-finding method based on the well-known friends-of-friends algorithm, see \cite{davis}. \cite{tago2005} 
identified 7657 and 10058 groups of galaxies in the Northern (N) and Sourthern (S) parts of the 
2dF GRS, respectively.

These numbers are quite large compared with the numbers reported 
in this paper, because we obtained 1363 groups of gas clumps in the P7t cubic partition, see the right 
bottom panel of Fig.\ref{Fig:PropGru_Contf}. Nevertheless, some physical 
properties obtained here are similar to those reported by \cite{tago2005}. For instance, the size of most of the 
groups of galaxies detected by \cite{tago2005} is within 0.1-0.5 h$^{-1}\,$Mpc ( see the left panel of 
their Fig. 3). Meanwhile, the average effective radius of groups of galaxies found by \cite{tago2006} is 
0.61 h$^{-1}\,$Mpc. In this paper, we obtained an average group radius  
in the range of 0.5-0.8 Mpc, as can be seen in the right bottom panel of Fig.\ref{Fig:PropGru_Rad}. The dispersion 
of velocity found by \cite{tago2005} is similar for both the N and S parts of the 2dF GRS, around 200 km/s for 
groups of galaxies with less than 10 members ( see the left panel of 
their Fig. 4). In this paper, we obtained an average dispersion of 
velocity within the range of 80 to 200 km/s, as can be seen in the right bottom panel of 
Fig.\ref{Fig:PropGru_SigmaVel}.

The mass of the groups of galaxies can be estimated by using the virial theorem and 
a typical size and velocity dispersion. For groups with three or a very few more member 
galaxies, this dynamical mass is within the 
range 10$^{12.5}$ to 10$^{14}$ M$_{\odot}$, see \cite{geller}, \cite{nol}, 
\cite{eke2}, \cite{ber}, \cite{yang} and \cite{yang2}. In addition, by using a large set of galaxies 
contained in the fourth data release of the SDSS-DR4, see \cite{adelman},  
\cite{zandivarez} cataloged groups of galaxies, whose relationship between environment and physical properties was studied by 
\cite{martinez}. In the bottom right-hand panel of Fig.1, \cite{martinez} reported the log of 
the mass distribution of the groups sample against the number of groups with more than four members. They found that most of 
the groups have a log(M) around 13, so that it is in good agreement with tha mass reported in the right-hand panels of our 
Fig.\ref{Fig:PropGru_Masa} for gas groups of multiplicity 4.          
         
%%%%%%%%%%%%%%%%%%%%%%%%%%%%%%%%%%%%%%%%%%%%%%%%%%%%%%%%%%%%%%%%%%%%%%%%%%%%%%%%%%%%%%%%%%%%%%%%%%%%%%%%%%%
\section{Concluding Remarks}
\label{sec:conclu}

Some galaxy surveys, as the Euclid Space Mision ( see \cite{lau} ), the Sloan Digital Sky Survey ( see \cite{york} ), the 
Plank Survey ( see \cite{planck2005}, \cite{planck2014} and \cite{planck2016}) , among 
others,  will deliver new data shortly, so that the properties of galaxy clustering and galaxy groups 
will continue to be probes to study the growth of large scale structure of the Universe. For this reason, the 
development of algorithms to detect groups and cluster of galaxies is always needed.   

With this purpose in mind, in this paper several uniform cubic partitions of the simulation volume were 
implemented to detect isolated dense gas clumps and calculate their clustering and grouping properties 
at galaxy cluster scales in a typical cosmological simulation.

Throughout Sections \ref{subs:funcionmasa} to 
\ref{subs:grupos}, and particularly in the first paragraphs of 
Section \ref{sec:discu}, we demonstrated that the low-resolution partitions 
P6 and P6t do not have enough resolution to describe the clustering and grouping properties considered, 
while the higher-resolution partitions P7 and P7t do a better representation. A higher level 
partition applied to a more resolved simulation will give better results at a much higher computational cost. 

It is less conclusive is if there is any advantage when changing the normalized density threshold 
for a couple of partitions of the same resolution, such as P6 to P6t or P7 to P7t. We remark that 
partition P7t detected the largest multiplicity group of the simulation, even when compared to the partition P7. 

In addition, we note that the highest multiplicity groups presented in section \ref{subs:grupos}, do not generally 
have the largest size, mass and velocity dispersion when compared with lower multiplicity groups detected with 
the same partition. It can be expected that these physical properties of a group 
increase with respect to its multiplicity because more gas cloud members will need more space, will 
accumulate more mass and its further away particles will go faster. 

Therefore, we conclude that our description of the grouping properties based on cubic partitions 
is partially acceptable because the low-multiplicity groups detected in this paper are in better agreement with 
these physical expectations.

In the simulation presented in this paper, the structures detected in small groups 
have a geometrical size within the range of 0.45--1.25 Mpc (mega-parsec); as discussed in 
Section \ref{subs:grupos}. It is likely that most of the dense gas clumps forming these small groups 
will continue to collapse gravitationally, so that their size 
will diminish while their density will increase up to the point where galaxies be formed. It is therefore possible 
that the dynamics of the well-observed small groups of galaxies at kpc scale have been partially inherited from 
groups formed at a larger spatial scale of Mpc. The most important prediction of this work consists of the fitted curves that 
indicate how the physical properties of the groups will change in terms of the number 
of members or multiplicity, see Table \ref{tab:fitingf}. 

To establish the goodness of the fitting models, the $\chi_{\nu}^{2}$ statistical test has been applied 
to the data. In principle, the two fits for the number of groups, $N_g$ and for the velocity dispersion $\sigma_v$ would qualify 
as poor fitting models; the fit for the log of the mass and the radius seem to be "over-fitted". For this reason, new fitting 
models need to be proposed. 

To continue with the inspection of the certainty of the parameters of the fitting models, the confidence interval 
has been calculated and shown in Table \ref{tab:confidence}. The confidence intervals are very wide in general, except 
for the fitting parameters of the log of the mass $< \log \left( M / M_{\odot}\right)> $ and the radius $<r>$.

To take advantage of all the physical information displayed at the vertical axis of 
Figures \ref{Fig:PropGru_Contf}-\ref{Fig:PropGru_SigmaVel} and 
at the same time, to compare the results with the two sets of grids P7 and P6, to assess the influence of the grid on the study 
undertaken in this paper, in Tables \ref{tab:confidencephysicalP7} and \ref{tab:confidencephysicalP6} we showed the 
mean, the standard deviation and the confidence interval of each physical property.

By using the grid P6, a smaller number of low-multiplicity groups are detected, with a similar mass but with 
a slightly larger size and velocity dispersion in comparison with the properties 
obtained by using the grid P7. Being the log of the mass and radius the only physical properties which
seem to be grid-independent, then these results would allow us to speak about the average physical
properties expected for low-multiplicity groups of protogalaxies.

Thus, one can set forth the second prediction of this paper, which is that for low-multiplicity
groups of protogalaxies, the average log of the mass is within the interval 12.65-13.23 and the average
radius is within the interval 0.5-1.16 Mpc.

The third prediction in this paper is based on the calculation of the distribution of the chosen 
cubic elements in terms of their ratio values of $\frac{E_{\rm kin}}{\left|E_{\rm grav}\right|}$, which was shown in 
Fig.\ref{fig:FuncBeta}. It should be emphasized that we found that most of the gas clumps have a low-value 
of $\frac{E_{\rm kin}}{\left|E_{\rm grav}\right|}$. This result is in agreement with observational estimates 
of the ratio  $\frac{E_{\rm kin}}{\left|E_{\rm grav}\right|}$ in pre-stellar gas cores which are within the range 
10$^{-4}$-0.07, see \cite{caselli} and \cite{jijina}. 

Among the advantages of our method, the programming is easy and it leads 
in a natural way to the zoom-in technique, in which a cell of the partition can be chosen to re-simulate its 
content of matter and increase its number of particles to represent only that 
particular region in a new simulation box. 

The zoom-in technique has allowed a single chosen 
galaxy or dark matter halo to be studied in great detail, see for instance \cite{roca}. In our case, the 
results of this paper can be used as suitable initial conditions 
for numerical simulations that aim to follow the dynamical evolution of small groups of galaxies; for instance, 
\cite{barnes85} and \cite{aceves} have started their simulations with a value 
of $\frac{E_{\rm kin}}{\left|E_{\rm grav}\right|}$ in the range 0.1--1. 
    
%%%%%%%%%%%%%%%%%%%%%%%%%%%%%%%%%%%%%%%%%%%%%%%%%%%%%%%%%%%%%%%%%%%%%%%%%%%%%%%%%%%%%%%%%%%%%%%%%%%%%%%%%%%%%%%%%%%%%%%%
\section{acknowledgements}
The author gratefully acknowledges the computer resources, technical expertise, and
support provided by the Laboratorio Nacional de Superc\'omputo del Sureste de M\'exico
through grant number O-2016/047.
%%%%%%%%%%%%%%%%%%%%%%%%%%%%%%%%%%%%%%%%%%%%%%%%%%%%%%%%%%%%%%%%%%%%%%%%%%%%%%%%%%%%%%%%%%%%%%%%%%%%%%%%%%%%%%%%%%%%%%%%

%%%%%%%%%%%%%%%%%%%%%%%%%%%%%%%%%%%%%%%%%%%%%%%%%%%%%%%%%%%%%%%%%%%%%
%%%%%%%%%%%%%%%%%%%%%%%%%%%%%%%%%%%%%%%%%%%%%%%%%%%%%%%%%%%%%%%%%%%%%
\end{document}